\documentclass[twocolumn]{aastex6}

\bibpunct{(}{)}{;}{a}{}{,}
\begin{document}

\title{The impact of {\em JWST} broad-band filter choice on photometric redshift estimation}
\author{L. Bisigello\altaffilmark{1,2}, K. I. Caputi\altaffilmark{1}, L. Colina\altaffilmark{3}, O. Le F\`evre\altaffilmark{4}, H. U. N\o rgaard-Nielsen\altaffilmark{5}, P. G. P\'erez-Gonz\'alez\altaffilmark{6}, J. Pye\altaffilmark{7}, P. van der Werf\altaffilmark{8}, O. Ilbert\altaffilmark{4}, N. Grogin\altaffilmark{9}, A. Koekemoer\altaffilmark{9}
}

\altaffiltext{1}{Kapteyn Astronomical Institute, University of Groningen, P.O. Box 800, 9700 AV, Groningen, The Netherlands.}
\altaffiltext{2}{SRON Netherlands Institute for Space Research, 9747 AD, Groningen, The Netherlands. 
\\ Email:  bisigello@astro.rug.nl}
\altaffiltext{3}{Centro de Astrobiolog\'{\i}a, Departamento de Astrof\'{\i}sica, CSIC-INTA, Cra. de Ajalvir km.4, 28850 - Torrej\'on de Ardoz, Madrid, Spain}
\altaffiltext{4}{Aix Marseille Universit\'e, CNRS, LAM (Laboratoire d'Astrophysique de Marseille), UMR 7326, 13388, Marseille, France}
\altaffiltext{5}{National Space Institute (DTU Space), Technical University of Denmark, Elektrovej, DK-2800 Kgs. Lyngby, Denmark}
\altaffiltext{6}{Departamento de Astrof\'{\i}sica, Facultad de CC. F\'{\i}sicas, Universidad Complutense de Madrid, E-28040 Madrid, Spain; Associate Astronomer at Steward Observatory, The University of Arizona.}
\altaffiltext{7}{University of Leicester, Department of Physics \& Astronomy, Leicester, LE1 7RH, UK}
\altaffiltext{8}{Sterrewacht Leiden, Leiden University, PO Box 9513, 2300 RA, Leiden, The Netherlands}
\altaffiltext{9}{Space Telescope Science Institute, 3700 San Martin Drive, Baltimore, MD 21218, USA}

\shorttitle{Bisigello et al.: Photometric redshifts with {\em JWST} filters}

\shortauthors{Bisigello et al.}

\begin{abstract}

The determination of galaxy redshifts in {\em James Webb Space Telescope (JWST)'s} blank-field surveys will mostly rely on photometric estimates, based on the data provided by {\em JWST's} Near-Infrared Camera (NIRCam) at $0.6-5.0 \, \rm \mu m$ and Mid Infrared Instrument (MIRI) at $\lambda>5.0 \, \rm \mu m$.  In this work we analyse the impact of choosing different combinations of NIRCam and MIRI broad-band filters (F070W to  F770W), as well as having ancillary data at $\lambda < 0.6 \, \rm \mu m$, on the derived photometric redshifts ($z_{\rm phot}$) of a total of 5921 real and simulated galaxies, with known input redshifts $z=0-10$. We found that observations at $\lambda<0.6 \, \rm \mu m$ are necessary to control the contamination of high-$z$ samples by low-$z$ interlopers. Adding MIRI  (F560W and F770W) photometry to the NIRCam data  mitigates the absence of ancillary observations at $\lambda<0.6 \, \rm \mu m$ and improves the redshift estimation. At $z=7-10$, accurate $z_{\rm phot}$ can be obtained with the NIRCam broad bands alone  when $S/N \geq 10$, but the $z_{\rm phot}$ quality significantly degrades at $S/N \leq 5$.  Adding MIRI photometry with one magnitude brighter depth than the NIRCam depth allows for a redshift recovery of $83-99\%$, depending on SED type, and its effect is particularly noteworthy for galaxies with nebular emission. The vast majority of NIRCam galaxies with [F150W]=29~AB~mag at $z=7-10$ will be detected with MIRI at [F560W, F770W]$<28$~mag if these sources are at least mildly evolved or have spectra with emission lines boosting the mid-infrared fluxes.

\end{abstract}


\section{Introduction} \label{sec:intro}
The power of multi-wavelength photometric observations to recover galaxy redshifts has been known since the late fifties \citep{Baum1957} and it has been confirmed in the last decade with deep blank-field imaging surveys \citep[e.g.][]{Ilbert2009,Cardamone2010,Dahlen2013, Rafelski2015}. Although spectroscopic redshifts are much better in precision than photometric estimates, photometry has two important advantages with respect to spectra: the number of sources that can be observed at the same time and sensitivity. Indeed, with a single photometric pointing it is possible to observe a large number of sources, while in general only a limited number of objects can simultaneously be targeted with spectroscopy. In addition, the spectroscopic observation of faint targets usually requires very long exposure times \citep[e.g.][]{Lefevre2015}, and the faintest objects detected in photometric maps are beyond the technical possibilities of contemporary spectrographs \citep{Caputi2012,Caputi2015}.\par

Photometric redshift determinations usually rely on the identification of strong features, such as the Lyman or $4000 \, \rm \AA$ break, in a galaxy spectral energy distribution (SED), after they are convolved with the transmission functions of the filters utilised in the observations. This is the reason why it is necessary to do a careful filter selection when planning observations, balancing depth and wavelength coverage, in order to minimise degeneracies and misidentifications when obtaining photometric redshifts. 
\par

The {\em James Webb Space Telescope} ({\em JWST}\footnote{http://www.jwst.nasa.gov}, \citealt{Gardner2009}) is a foremost space mission for the coming years and the awaited successor of the {\em Hubble Space Telescope (HST)} and  {\em Spitzer Space Telescope} at infrared wavelengths. It will have four instruments on board for imaging, spectroscopy and coronography, covering a  wide range of wavelengths from 0.6~$\rm \mu m$ through 28.3~$\rm \mu m$ with sub-arcsec angular resolution. \par

One of {\em JWST'}s main scientific aims is to study the formation and evolution of galaxies at early cosmic times. For this purpose, deep blank-field imaging surveys will be carried out with two imaging cameras, namely the Near Infrared Camera  \citep[NIRCam;][]{Rieke2005} and the Mid Infrared Instrument \citep[MIRI;][]{Rieke2015,Wright2015}. In the vast majority of cases, galaxy redshift determinations will be done through SED-fitting analysis and, therefore, it is crucial to understand the impact of choosing different filter combinations on the ability to recover the right redshifts for all the observed sources.  \par

NIRCam's filter wheels include eight broad-band filters, namely F070W, F090W, F115W, F150W, F200W, F277W, F356W and F444W, as well as a number of medium and narrow-band filters, covering the wavelength range $0.6-5.0 \, \rm \mu m$\footnote{http://www.stsci.edu/jwst/instruments/nircam}. The MIRI imager\footnote{http://www.stsci.edu/jwst/instruments/miri}  contains nine different broad-band filters between 5 and 28~$\mu$m.  This latter wavelength range encompasses the redshifted stellar light of high-$z$ galaxies, as well as the hot dust radiation emitted by low-$z$ sources. The two shortest-wavelength bands that we consider here (i.e., F560W and F770W) are the most sensitive in MIRI, so will be those primarily used for high-$z$ galaxy studies.\par

The complementarity of NIRCam's and MIRI's wavelength coverages makes that both cameras should ideally be considered for deep galaxy surveys. While surveys with NIRCam alone will be sufficient to encompass the $4000 \, \rm \AA$ break of galaxies at $0.6 <z < 10$, a secure measurement of the evolved stellar light (around rest $1 \, \rm \mu m$) in $z>4$ galaxies will also require observations with the MIRI  filters. However, due to different detector technology in the near- and mid-IR range, MIRI's sensitivity is significantly lower than that of NIRCam, making observers doubt whether it is convenient to invest time on long MIRI observations when planning deep galaxy surveys. At the same time, the lack of {\em JWST} coverage below $0.6 \, \rm \mu m$ requires considering whether observing fields with no ancillary shorter wavelength imaging can be effective to study galaxy evolution. Here we address these issues through the analysis of galaxy redshift recovery in different data availability situations, all treated on an equal basis, in order to allow the reader decide which camera and filter configuration would be optimal for their scientific interests.

The aim of this work is to assess the impact of different {\em JWST} broad-band filter combinations on recovering reliable photometric redshifts for galaxies at $z=0-10$. In addition to considering the NIRCam broad bands and MIRI F560W and F770W filters, we tested the need of ancillary data at $\lambda<0.6 \, \rm \mu m$, such as those provided by {\em HST} observations. We applied our tests to three different galaxy samples with known redshifts: 1) a spectroscopic galaxy sample at $z=0-6$; 2) a galaxy sample with consensus photometric redshifts at $z=4-7$ from the CANDELS survey; and 3) a mock galaxy sample generated with different spectral templates at $z=7$, 8, 9 and 10. \par

The structure of the paper is as follows. In section \ref{sec:samples} we describe our three sample selection, the obtention of photometry in the relevant NIRCam and MIRI bands,  and photometric redshift test methodology.  In section \ref{sec:results} we analyse our results at different redshifts, while in section \ref{sec:conclusions} we summarize our main findings and conclusions. Throughout this paper, we adopt a cosmology with $H_0=$70~$\rm km \, s^{-1} \, Mpc^{-1}$, $\Omega_M=0.27$, $\Omega_\Lambda=0.73$. All magnitudes refer to the AB system \citep{Oke1983}. \par

\section{Sample selection and test methodology}\label{sec:samples}

\subsection{Galaxy sample selection}

We selected three different galaxy samples spanning rather complementary redshift ranges: 
\begin{enumerate}
  \item a sample of 2422 galaxies with secure spectroscopic redshifts at $z=0-6$;
  \item a sample of 1375 galaxies with consensus photometric redshifts at $z=4-7$;
  \item a sample of 2124 simulated galaxies at $z=7-10$. 
\end{enumerate}
\par

These three samples allowed us to investigate the different problems arising when trying to recover redshifts photometrically using different combinations of {\em JWST} filters. By no means are these samples trying to emulate a complete, flux-limited galaxy population at $z=0-10$ as it will be seen in a blank {\em JWST} field, and this is not necessary given the scope of this paper.

\subsubsection{Sample 1}
\label{sec-sample1}

Our first galaxy sample contains 2422 galaxies with secure spectroscopic redshifts ($z_{\rm spec}$) from the ESO public compilation in the Chandra Deep Field South \citep[CDFS;][]{Cristiani2000,Croom2001,Bunker2003,Dickinson2004,Stanway2004,Stanway2004b,Strolger2004,Szokoly2004,vanderWel2004,Doherty2005,Lefevre2005,Mignoli2005,Ravikumar2007,Popesso2009,Balestra2010,Silverman2010,Kurk2013,Vanzella2014}, and from \citet{Morris2015}. In order to obtain multiwavelength photometry for these galaxies, we crossmatched this sample with the public Cosmic Assembly Near-infrared Deep Extragalactic Legacy Survey \citep[CANDELS;][]{Grogin2011,Koekemoer2011} catalogue for the Great Observatory Origins Deep Survey South field (GOODS-S) obtained by \citet{Guo2013}. This catalogue contains photometry in 17 broad bands, from 0.37 $\mu$m through 8.0 $\mu$m, obtained with ground-based  and space telescopes, including {\em HST} and {\em Spitzer} \citep{Ashby2015}. This wavelength sampling allows for a good-quality SED fitting of the galaxy stellar emission and a proper redshift recovery in the vast majority of cases, using standard galaxy spectral templates and photometric redshift codes.\par

Our aim here is to assess the impact of choosing different {\em JWST} filter sets on the derived photometric redshifts, rather than the performance of different photometric redshift codes and/or SED template libraries. Therefore,  we explicitly excluded from our analysis any galaxy from the parent spectroscopic sample for which we could not correctly recover the redshift photometrically using all the CANDELS photometry up to $4.5 \, \rm \mu m$.  We did not use the two IRAC band data at longer wavelengths because their $S/N$ is considerably lower than for the other bands and they could be contaminated by polycyclic aromatic hydrocarbon emission at low $z$. The excluded galaxies are only $\sim 8\%$ of all the galaxies with publicly available $z_{\rm spec}$ in the CDFS. \par

To obtain photometric redshift estimates for all the galaxies in the parent spectroscopic sample,  we ran the public code \textit{LePhare} \citep{Arnouts1999,Ilbert2006} using templates from Bruzual \& Charlot~(2003; BC03 hereafter) with solar metallicity,  a range of exponentially declining star formation histories with different characteristic times $\tau$ from 0.01 to 10~Gyr and ages from 0.01 to 10~Gyr. We allowed for emission lines in \textit{LePhare}, and applied the Calzetti et al. reddening law \citep{Calzetti2000} with $A_{V}$ extinction values from 0 to 4. \par

\begin{figure}[ht!]
\center{
\includegraphics[trim={0cm 0cm 0cm 0cm},clip,width=1\linewidth, keepaspectratio]{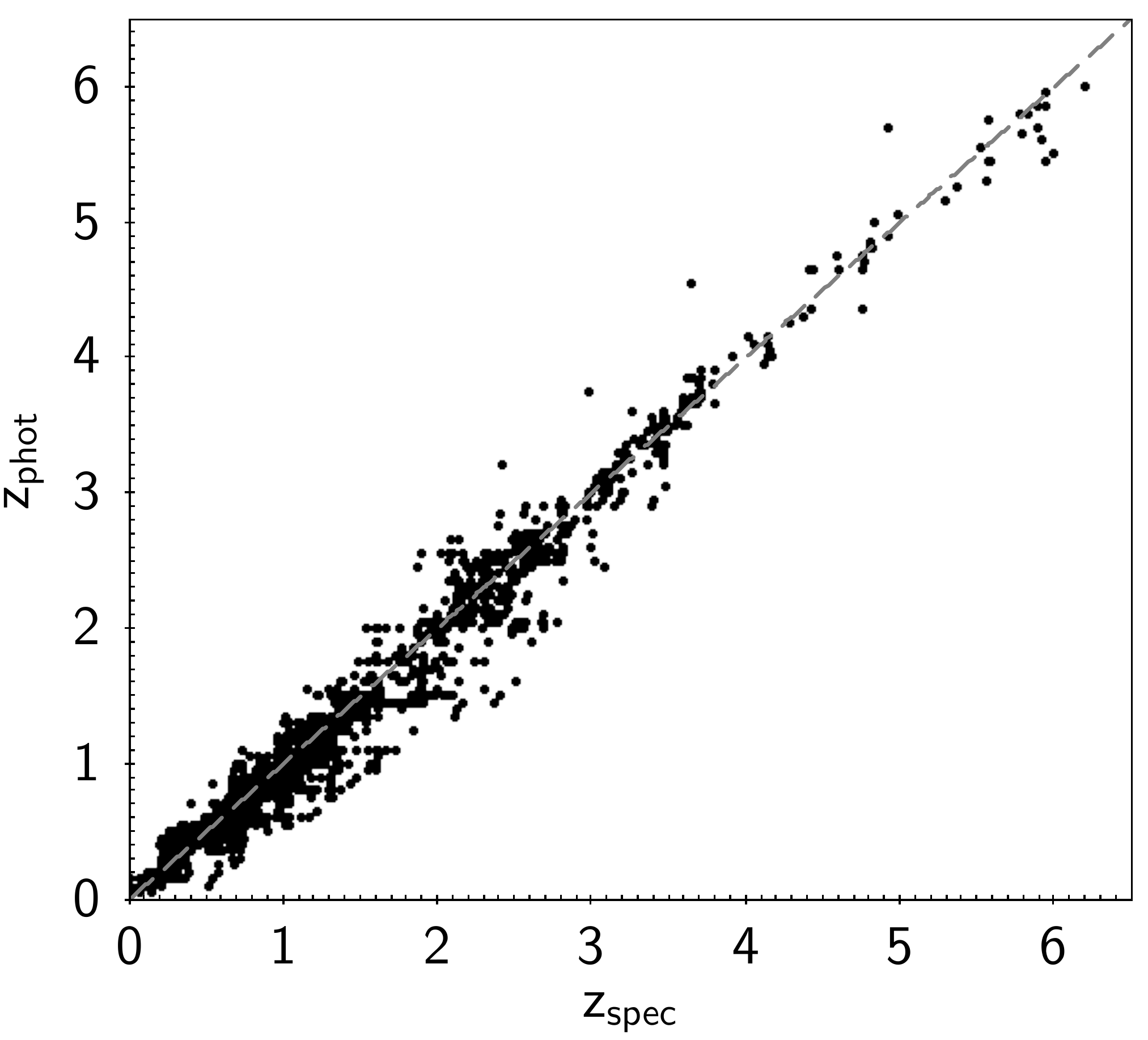}}
\caption{Photometric redshifts obtained with CANDELS 15-band photometry versus spectroscopic redshifts for Sample 1.\label{fig:zspec}}
\end{figure}

\begin{figure}[ht!]
\center{
\includegraphics[width=1\linewidth, keepaspectratio]{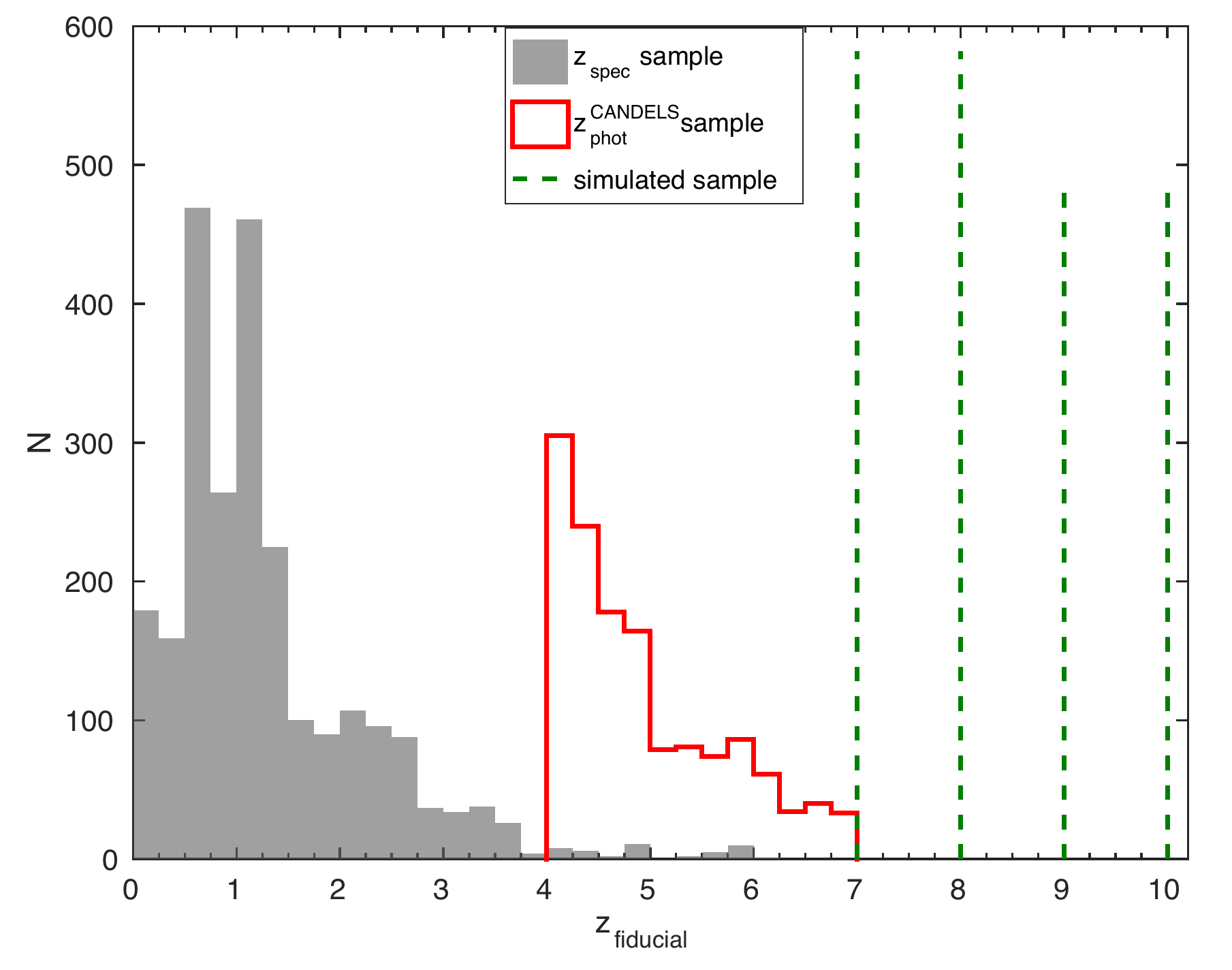}}
\caption{Redshift distribution of the three galaxy samples considered in this work: a galaxy sample with spectroscopic redshifts in the GOODS-S field (full grey; Sample 1), a galaxy sample with well established CANDELS photometric redshifts, also in GOODS-S (red line; Sample 2) and a sample of simulated galaxies with known input redshifts (dashed line; Sample 3).  \label{fig:zdist}}
\end{figure}

We considered the subsample of 2422 objects with resulting photometric redshifts within 2$\sigma$ ($\sigma=$0.13) of the spectroscopic redshifts (Fig. \ref{fig:zspec}), which constitute 92\% of the parent spectroscopic sample, and adopted it as our  Sample 1 for all further analysis. The Sample 1 redshift distribution is shown in Figure \ref{fig:zdist}.  This distribution peaks around $z\sim1$ and has a high-$z$ tail up to $z=6$, although the statistics are rather poor at $z>4$.  So, our main purpose with Sample 1 is to study the percentage of low-redshift galaxies that can leak to high $z$ when different filter sets, and thus different wavelength coverages, are available. \par


\subsubsection{Sample 2}

Our second sample consists of 1375 galaxies at $z=4-7$ taken from the CANDELS GOODS-S public catalogue. For this sample, we considered as fiducial redshifts the consensus photometric redshifts that have been compiled by \citet{Dahlen2013}. The CANDELS consensus photometric redshifts result from the comparison of 11 independent SED fitting runs. We considered only CANDELS galaxies with $z>4$ here, as lower redshifts are well represented in our Sample 1.

\begin{figure}[ht!]
\center{
\includegraphics[trim={0cm 0cm 0cm 0cm},clip,width=1\linewidth, keepaspectratio]{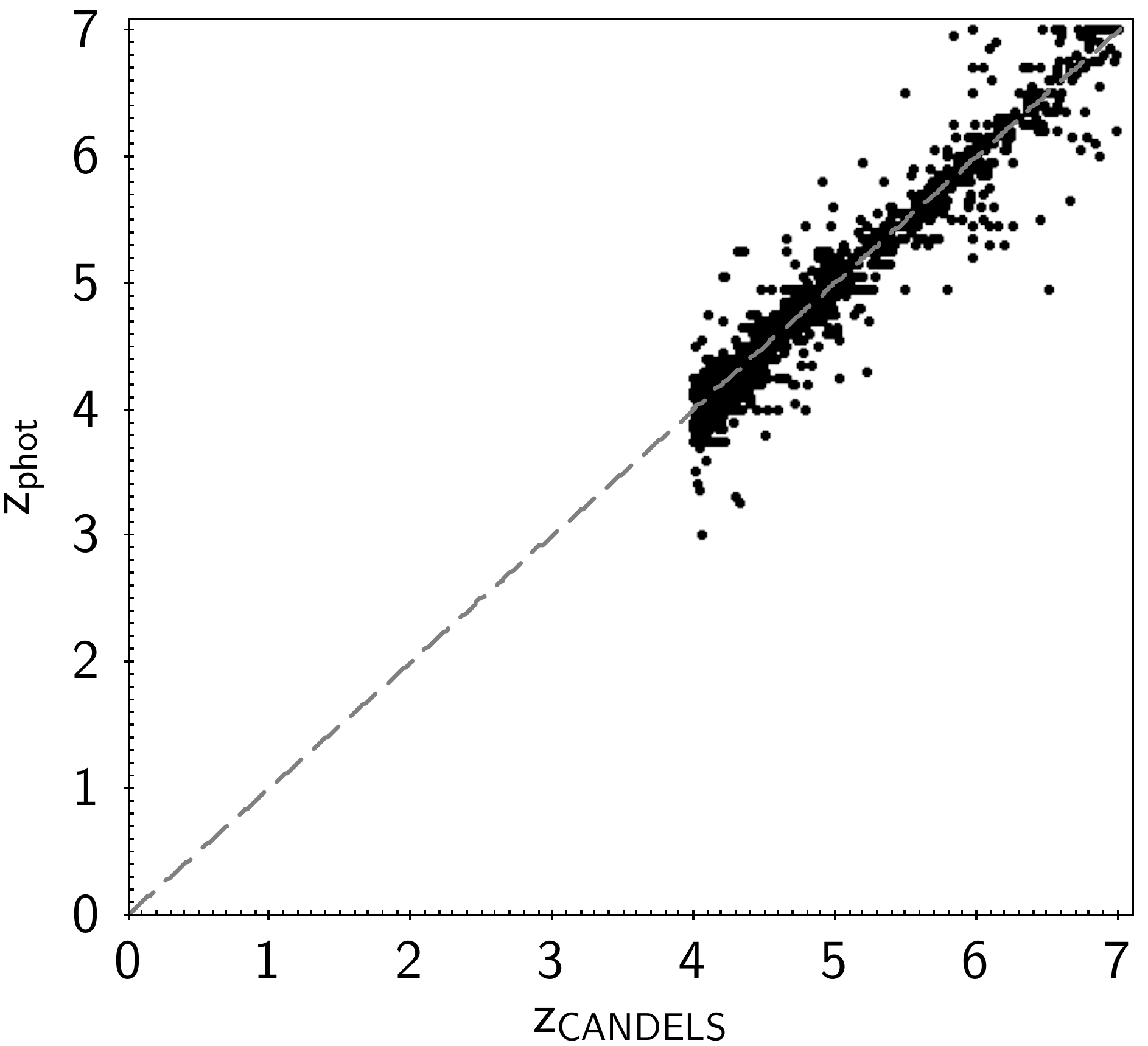}}
\caption{Photometric redshifts obtained with 15-band CANDELS photometry in this work versus the consensus CANDELS photometric redshifts for Sample 2. \label{fig:zphot}}
\end{figure}

As we did for the sample with spectroscopic redshifts (Sample 1), we obtained our own photometric redshifts by running \textit{LePhare} on the CANDELS 15-band photometry and using the same SED templates and parameter values stated before. In this case, our own photometric redshifts agree within 2$\sigma$ ($\sigma=$0.10) with the CANDELS consensus redshifts for 1375 (95\%) of the GOODS-S galaxies with CANDELS redshifts $z_{\rm CANDELS}=4-7$ (Fig. \ref{fig:zphot}), consistently with the typical level of dispersion present in the individual SED fitting CANDELS runs. We adopted these 1375 galaxies as our Sample 2.  The resulting redshift distribution is shown in Figure \ref{fig:zdist}. \par

\subsubsection{Sample 3}
\label{sec_sample3descr}

We built our third sample with mock galaxies at fixed redshifts z$=7, 8, 9$ and $10$ (for an analysis of more redshifts see Appendix \ref{sec:AppA}), simulating their SEDs with two different methods. Firstly, we considered the BC03 synthetic galaxy templates with four metallicities (Z$_{\odot}$, 0.4~Z$_{\odot}$, 0.2~Z$_{\odot}$, and 0.02~Z$_{\odot}$), exponentially declining star formation histories and five different ages, depending on redshift. We applied dust extinction by using the Calzetti et al. reddening law with $A_{V}=0-1$, and  the effect of absorption by the intergalactic medium \citep[IGM;][]{Madau1995}. 

The BC03 models do not include emission lines, so we manually added the $\rm H\alpha$, $\rm H\beta$, [OIII]5007~$\rm \AA$ and [OII]3727~$\rm \AA$ emission lines to those templates for which the age is equal or lower than the characterisitic $\tau$ of the exponentially declining star formation history, as follows. We started from the observed mean value of the $H\alpha+$[NII]$+$[SII]  equivalent width (EW), which is 422-423~${\rm \AA}$ at $z=3.8-5$ \citep{Smit2015}. Then, we assumed line ratios from \citet{Anders2003} in order to derive the equivalent widths of $H\alpha$ alone at different metallicities. \citet{Fumagalli2012} obtained a relation between $\rm H\alpha$ equivalent width and redshift, namely EW$\rm (H\alpha)$~$\propto (1+z)^{1.8}$, which has been checked to be consistent with observations up to $z\sim5$ \citep{Stark2013} \footnote{This redshift dependence is slightly steeper than that obtained by e.g. Faisst et al.~(2016), but adopting this other prescription would only change our assumed H$\alpha$ EW by a factor $<1.3$.}. Following this relation, we derived EW($\rm H\alpha$) for all considered redshifts. We also derived equivalent widths for $\rm H\beta$ assuming case B recombination and the corresponding equivalent widths for [OIII] and [OII] by using the line ratios tabulated in \citet{Anders2003}. We explicitly decided not to include the Ly$\rm \alpha$ line because the importance of its emission in galaxies at $z>7$ is still unclear \citep[e.g.,][]{Stark2013,Konno2014,Pentericci2014}. Finally, we normalised every model at 29~AB~mag at 1.5$\mu$m (corresponding to the NIRCam F150W filter pivot wavelength). 

\begin{figure}[ht!]
\center{
\includegraphics[width=1\linewidth, keepaspectratio]{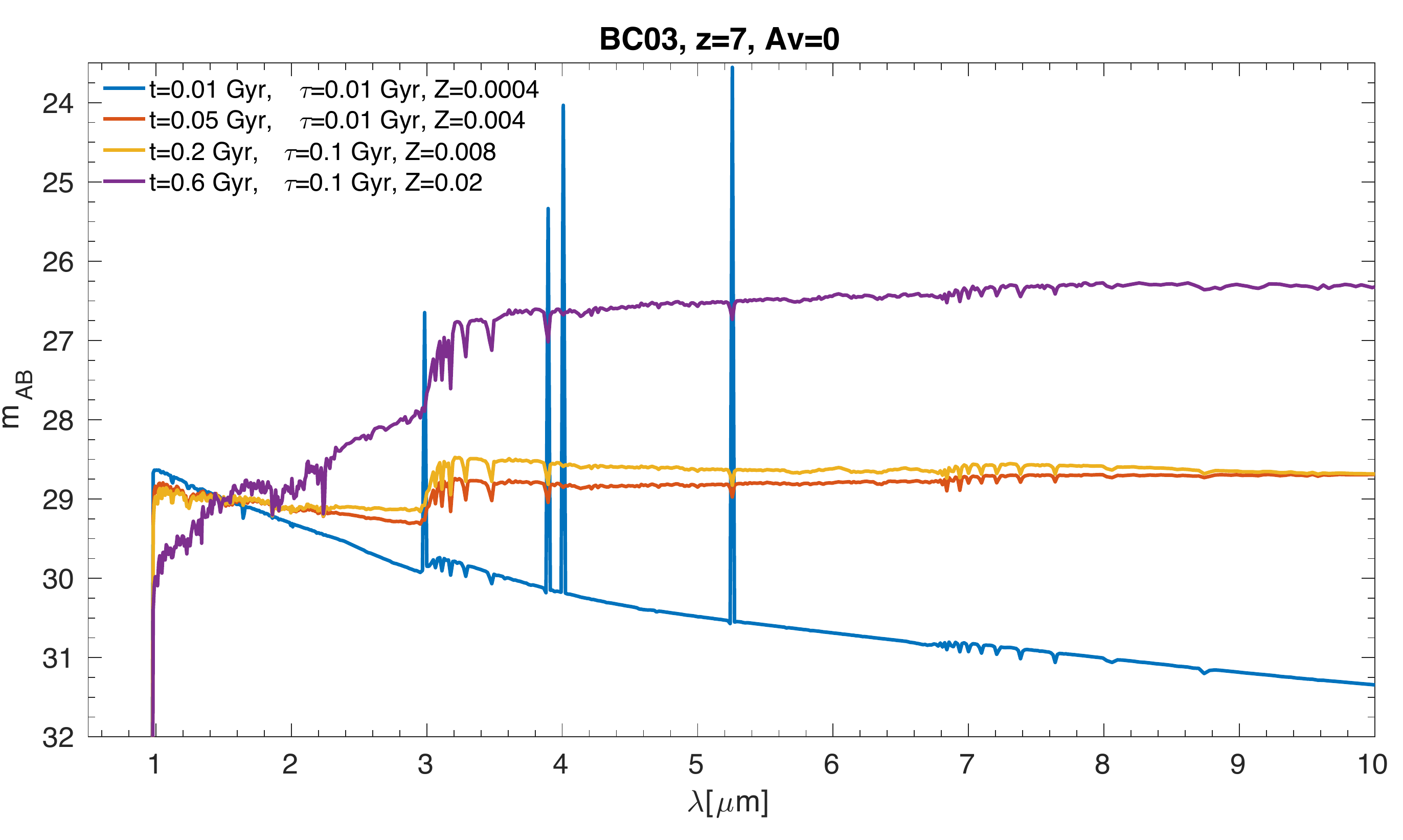}}
\caption{Four examples of BC03 SED templates at z$=$7 with $A_V=0$.  \label{fig:sed_z7}}
\end{figure}

The complete list of adopted parameter values is summarized in Table \ref{tab:param}, while some SED examples at z$=$7 are shown in Figure \ref{fig:sed_z7}. Our sample contains a total of 864 galaxies, corresponding to 240 BC03-based SED models at each redshift 7 and 8, and 192 SED models at each redshift 9 and 10.  For each model, we considered independently three possible $S/N$ values ($S/N=3, 5$ and 10) for the flux corresponding to [F150W]=29~AB~mag.  The $S/N$ values of the photometry in all other bands have been scaled as we explain in \S\ref{sec-jwstphotmeas}.  \par

\begin{deluxetable}{cc}[]
\tablecaption{Parameter values used to create BC03 SED models for simulated galaxies at z$=7-10$.  \label{tab:param}}
\tablecolumns{2}
\tablenum{1}
\tablewidth{0pt}
\tablehead{
\colhead{Parameter} &
\colhead{Values}
}
\startdata
metallicity & Z$_{\odot}$, 0.4Z$_{\odot}$, 0.2Z$_{\odot}$,0.02Z$_{\odot}\tablenotemark{a}$\\
SFH $\tau$  [Gyr] & 0.01,0.1,1,10 \\
A$_{V}$ & 0,0.5,1 \\
age [Gyr] & 0.01,0.05,0.2,0.4,0.6\tablenotemark{b}\\
$z$ & 7,8,9,10\\
$S/N$\tablenotemark{c} & 3,5,10\\
\enddata
\tablenotetext{a}{for metallicity 0.02Z$_{\odot}$ we considered only ages $t<$0.2~Gyr.}
\tablenotetext{b}{we considered this age only up to redshift $z=8$.}
\tablenotetext{c}{for the flux corresponding to the normalization magnitude, i.e., [F150W]=29~mag.}
\end{deluxetable}

\begin{figure}[ht!]
\center{
\includegraphics[width=1\linewidth, keepaspectratio]{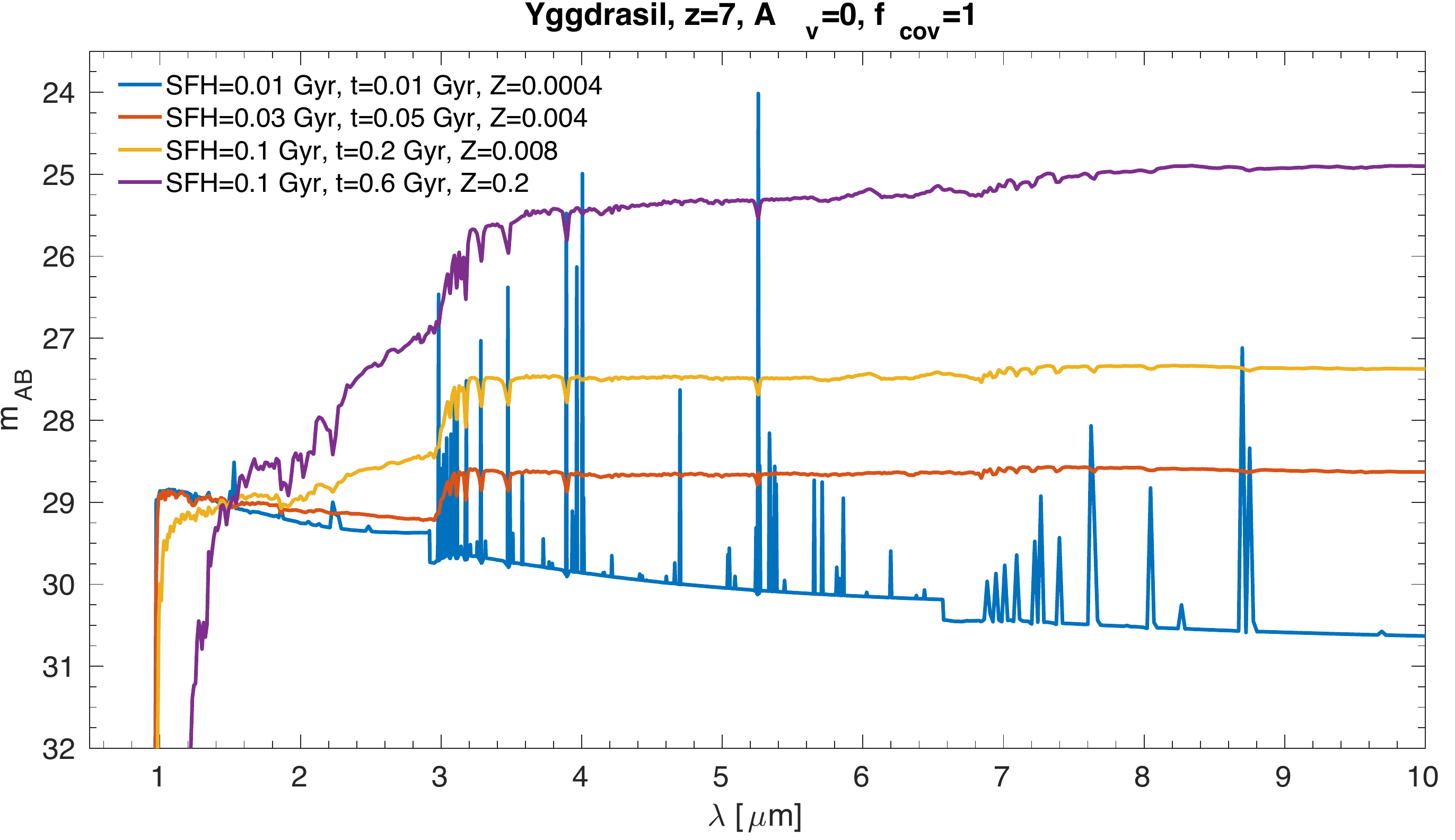}}
\caption{Four examples of {\em Yggdrasil} SED templates at $z=7$ with $A_V=0$. Note that templates with gas covering factor $f_{\rm cov}=1$ have multiple emission lines, but only when a young stellar population, due to very recent star formation, is still present. \label{fig:sed_z7_Ygg}}
\end{figure}

Secondly, we simulated further $z=7-10$ galaxy SEDs with the population synthesis code {\em Yggdrasil} \citep{Zackrisson2011}, which has been especially conceived to model high-redshift galaxies. We used a set of parameter values similar to the ones used for the BC03 templates, except that the star formation histories are somewhat different, consisting of a step function for which the duration of the constant star formation period is set as a free parameter. And, in addition to the normal stellar templates, we adopted also a template with a metal-free, population-III single stellar population \citep{Schaerer2002}. In all cases, we considered the two possible extreme gas covering factors, namely $f_{\rm cov}=0$ and 1. The complete list of adopted parameter values is given in Table \ref{tab:paramYgg}. We show four SED template examples for $z=7$ galaxies in Figure \ref{fig:sed_z7_Ygg}.

The {\em Yggdrasil} models include the effects of nebular line and continuum emission, so we did not need to add any extra line manually as we did for the BC03 templates. We did include the effect of dust reddening and IGM absorption in the same way as for the BC03 models. We removed the Ly$\alpha$ emission line from the {\em Yggdrasil} models, because its presence and importance is debated at $z>7$, as we explained before. Once again, we normalised every SED template to [F150W]=29~mag at 1.5~$\mu$m. Our {\em Yggdrasil} simulated sample comprises a total of 1260 galaxies, corresponding to 342 SED models at each redshift 7 and 8, and 288 SED models at each redshift 9 and 10.  The complete redshift distribution of our galaxy Sample 3, simulated  both with the BC03 and {\em Yggdrasil} templates, is shown in Fig. \ref{fig:zdist}.  \par

\begin{deluxetable}{cc}[ht!]
\tablecaption{Parameter values used to create {\em Yggdrasil} SED models for simulated galaxies at z$=7-10$. \label{tab:paramYgg}}
\tablecolumns{2}
\tablenum{2}
\tablewidth{0pt}
\tablehead{
\colhead{Parameter} &
\colhead{Values}
}
\startdata
metallicity & Z$_{\odot}$,0.4Z$_{\odot}$,0.2Z$_{\odot}$,0.02Z$_{\odot}$\tablenotemark{a},0\tablenotemark{a}\\
constant SFH for [Myr] & 10,30,100 \\
A$_{V}$ & 0,0.5,1 \\
age [Gyr] & 0.01,0.05,0.2,0.4,0.6\tablenotemark{b}\\
$f_{\rm cov}$ & 0,1\\
$z$ & 7,8,9,10\\
$S/N$\tablenotemark{c} & 3,5,10\\
\enddata
\tablenotetext{a}{for this metallicity we considered only ages $t<$0.2~Gyr.}
\tablenotetext{b}{we considered this age only up to redshift $z=8$.}
\tablenotetext{c}{for the flux corresponding to the normalisation magnitude, i.e., [F150W]=29~mag.}
\end{deluxetable}

\subsection{Test methodology}
\subsubsection{Interpolation/extrapolation of JWST photometry}
\label{sec-jwstphotmeas}

Based on the best-fit SEDs obtained running \textit{LePhare} on the CANDELS photometry for the galaxies in Samples 1 and 2, and the simulated SEDs for our mock galaxies in Sample 3,  we measured the photometry expected at the 8 NIRCam broad bands and the two MIRI broad bands F560W and F770W for all our galaxies (see Table \ref{tab:filters} for a summary of the different {\em JWST} filters considered in this work).  Our procedure to interpolate/extrapolate the {\em JWST} photometry is similar for our three galaxy samples. The main difference is that in Samples 1 and 2, we simply adopted $S/N$ values for the photometry scaled from the CANDELS photometry, while for Sample 3 we tested three different $S/N$ values in a controlled manner, in order to assess the effect of $S/N$ on our results. \par

As a first step, for every galaxy in our real galaxy samples (Samples 1 and 2), we derived NIRCam broad-band fluxes by convolving the best-fit SED template with the transmission curve of each {\em JWST} filter \citep{Meyer2004}. To each flux, we assigned an error bar corresponding to a $S/N$ equal to that of the closest filter in the CANDELS catalogue. We adopted as NIRCam flux a random value (selected from a uniform distribution) within the corresponding error bar. \par

Note that, in practice, NIRCam will reach the depth of the {\em HST} CANDELS photometry in very short integration times, so the $S/N$ achieved for CANDELS galaxies in typically deep NIRCam surveys will be higher than what we consider here. However, our assumed $S/N$  are already sufficiently high in most cases:  virtually all galaxies in Sample 1 and more than 90\% of galaxies in Sample 2 have $S/N \geq5$ in their derived NIRCam F150W photometry. Thus, our results derived here for these real galaxy samples are mostly independent of $S/N$, as we explain below  (the low $S/N$ effect will be more clearly manifested in our Sample 3, where we explicitly set three different $S/N$ values for each template, including a low value $S/N=3$). \par

As a second step, we measured the photometry expected at the MIRI bands for our Sample 1 and 2 galaxies, which results from the extrapolation of the best-fit template redward of the CANDELS IRAC $4.5 \, \rm \mu m$ coverage.  We convolved the best SED model with the two MIRI bands' transmission curves \citep{Bouchet2015,Glasse2015}. To each MIRI band flux, we  assigned an error bar corresponding to a $S/N$ scaled using as reference the $S/N$ in the NIRCam F150W band:  we assumed that a MIRI F560W or F770W flux corresponding to a magnitude 28 (AB) would have the same $S/N$ as a NIRCam F150W flux corresponding to 29~mag (whose $S/N$ was, in turn, scaled assuming that the `measured' NIRCam F150W flux $S/N$ for each galaxy was the same as the real CANDELS {\em HST} F160W flux $S/N$). Fixing the $S/N$ in the NIRCam and MIRI photometry implicitly assumes specific integration times with the {\em JWST} instruments.  In practice, this assumption does not matter, as our results will be equally valid for any galaxies with similar $S/N$ values, independently of the MIRI fluxes at which these $S/N$ are reached.\footnote{This statement implicitly assumes that the fainter galaxy SEDs will be represented by those of the currently analysed galaxies. So far, this has been shown to be the case with progressively deeper galaxy surveys.} Finally, we randomised every flux within its own error bar.\par

\begin{deluxetable}{ccc}[ht!]
\tablecaption{List of {\em JWST} filters considered in this work, their pivot wavelengths and bandwidths. \label{tab:filters}}
\tablecolumns{3}
\tablenum{3}
\tablewidth{0pt}
\tablehead{
\colhead{Filter name} &
\colhead{$\lambda$ [$\mu$m]} &
\colhead{Bandwidth [$\mu$m]\tablenotemark{a}}
}
\startdata
NIRCam/F070W & 0.70 & 0.18\\
NIRCam/F090W & 0.90 & 0.23\\
NIRCam/F115W & 1.15 & 0.30\\
NIRCam/F150W & 1.50 & 0.38\\
NIRCam/F200W & 1.99 & 0.53\\
NIRCam/F277W & 2.77 & 0.82\\
NIRCam/F356W & 3.56 & 0.99 \\
NIRCam/F444W & 4.41 & 1.29\\
MIRI/F560W & 5.60 & 1.41\\
MIRI/F770W & 7.70 & 2.30\\
\enddata
\tablenotetext{a}{This bandwidth corresponds to the entire wavelength range in which the filter transmission is $>1\%$.}
\end{deluxetable}

\begin{deluxetable*}{ccccccccccc}[ht!]
\tablecaption{List of 2$\sigma$ detection magnitude limits for different {\em JWST} filters corresponding to our fiducial source\tablenotemark{a} and each reference $S/N$ value considered for Sample 3. \label{tab:StoN_levels}}
\tablecolumns{11}
\tablenum{4}
\tablewidth{0pt}
\tablehead{
\colhead{Reference S/N\tablenotemark{a}} &
\colhead{F070W} &
\colhead{F090W} &
\colhead{F115W} &
\colhead{F150W} &
\colhead{F200W} &
\colhead{F277W} & 
\colhead{F356W} &
\colhead{F444W} &
\colhead{F560W} &
\colhead{F770W}
}
\startdata
10 & 30.2 & 30.5 & 30.6 & 30.7 & 30.9 & 30.9 & 30.7 & 30.2 & 29.7 & 29.7\\
5 & 29.4 & 29.7 & 29.9 & 30.0 & 30.2 & 30.2 & 30.0 & 29.4 & 29.0 & 29.0 \\
3 & 28.9 & 29.2 & 29.3 & 29.4 & 29.6 & 29.6 & 29.4 & 28.9 & 28.4 & 28.4 \\
\enddata
\tablenotetext{a}{$S/N$ value for [F150W]=29~mag.}
\end{deluxetable*}

For Sample 3 with mock galaxies, we convolved every simulated SED template  with the transmission curve of each NIRCam filter. In our sample every SED template is considered three times, in order to apply three fixed, reference $S/N$ values (10, 5 and 3) to the NIRCam F150W fluxes.
We assumed the same integration times for all the other NIRCam bands and took into account the differences in filter sensitivities to assign $S/N$ in each band. A different strategy that assumes the same observational depth for all NIRCam bands is presented in Appendix \ref{sec:AppB}. We considered every flux below the $2\sigma$ level as a non-detection. The 2$\sigma$ detection limits in each band for each fixed reference $S/N$ value at F150W are listed in Table~\ref{tab:StoN_levels}.   Note that, as a conservative approach, we adopted $3\sigma$ flux upper limits (rather than $2\sigma$) in all cases of non-detections for running the $z_{\rm phot}$ code, as it is explained below.

Finally, we obtained 10 random realisations of each NIRCam band flux within the corresponding error bar. Note that, in contrast, for Samples 1 and 2 we considered a single realisation of each flux, simply because in the case of real galaxies the NIRCam fluxes are well constrained by interpolation of the CANDELS fluxes, so their error bars are typically smaller than in Sample 3. 

In a similar way, we derived fluxes in the MIRI bands by convolving all SED templates in the sample with the two MIRI filter transmission curves. Then, we scaled each MIRI flux $S/N$, such that a mag$_{AB}=$28 in MIRI has the same $S/N$ as a NIRCam magnitude [F150W]=29. This is a realistic assumption of how the NIRCam and MIRI data will typically be matched. As for the NIRCam bands, we randomised every MIRI flux 10 times within its own error bar. \par

\subsection{Photometric redshift determinations}

With the expected {\em JWST} photometry measured on the galaxy templates for the three samples, we could test how accurately it is possible to derive photometric redshifts by using different {\em JWST} band combinations. The band combinations that we analysed in this work are:
\begin{itemize}
\item 8 NIRCam broad bands
\item {\em HST} F435W, F606W and 8 NIRCam broad bands
\item VLT $U$ band,  F435W, {\em HST} F606W and 8 NIRCam broad bands
\item 8 NIRCam broad bands and 2 MIRI bands (F560W and F770W)
\item 8 NIRCam broad bands and MIRI F560W only
\item 8 NIRCam broad bands and MIRI F770W only
\item {\em HST} F435W, F606W, 8 NIRCam broad bands and 2 MIRI bands
\item {\em HST} F435W, F606W, 8 NIRCam broad bands and MIRI F560W
\item {\em HST} F435W, F606W, 8 NIRCam broad bands and MIRI F770W
\end{itemize}

These different filter combinations have been chosen to test in which cases the NIRCam broad bands alone are sufficient to recover good photometric redshifts and in which cases the MIRI bands can improve these estimates. Because MIRI is considerably less sensitive than NIRCam, much longer integration times are necessary to reach comparable depths. Thus, many observing programs may opt for using a single MIRI filter (F560W or F770W) rather than both of them. This is why we tested separately the cases in which a single and both MIRI bands are used.

In addition, we analysed the cases in which two short-wavelength {\em HST} bands and the $U$ band are included. All these bands are beyond {\em JWST'}s wavelength coverage, so they cannot be substituted with any {\em JWST} band. We used the CANDELS catalogue to recover observations in these bands for galaxies in Samples 1 and 2. We ignored these bands for Sample 3: they are irrelevant, as the Lyman break is completely contained within the NIRCam wavelength range at $z\geq7$.

We note that \textit{LePhare} has been run using the native BC03 models allowing the code to incorporate emission lines with its own prescription in the cases of Samples 1 and 2 (as it is described in Section~\ref{sec-sample1}). For our mock galaxies in Sample 3,  we ran \textit{LePhare} using directly our own customised templates, i.e., the BC03 templates with emission lines incorporated as explained in Section~\ref{sec_sample3descr} and the \textit{Yggdrasil} templates. In addition, in the \textit{LePhare} runs for Sample 3 we also included some older galaxy models (as in Samples 1 and 2), and allowed for extinction values from $A_V=0$ to 4 in all cases, in order to test whether there could be degeneracies  between redshift and dust/age produced in the photometric fitting. This consideration of a wide range of redshifts, ages and extinction values emulates the real situation in which one simply has a photometric input catalogue without knowing a priori which sources are at high or low redshifts. No extra consideration of emission lines has been allowed for Sample 3 in \textit{LePhare}, as our young-galaxy models already account for them. \par

For all the filter combinations we ran \textit{LePhare} on the three galaxy samples, in order to recover the galaxy photometric redshifts and compare them with the fiducial redshift available for each galaxy. In the filter combinations with MIRI bands in Samples 1 and 2, and all cases for Sample 3, we chose the median photometric redshift of the 10 iterations done for each object (corresponding to different variations in the photometry, as explained above). 

For each sample and filter set combination, in Section~\ref{sec:results} we quote the resulting mean value of the normalised redshift difference distribution $\delta z=(z_{\mathrm{phot}}-z_{\mathrm{fiduc.}})/(1+z_{\mathrm{fiduc.}})$ and the r.m.s. ($\sigma$) of the {\em normalised} absolute redshift difference distribution  $|z_{\mathrm{phot}}-z_{\mathrm{fiduc.}}|/(1+z_{\mathrm{fiduc.}})$.  Note that, although we computed $\sigma$ independently for each sample and broad-band combination,  we considered as reference to define outliers a single $\sigma$ value per sample (the minimum obtained with all the filter sets).  This allows us to directly compare how the outlier fractions change among the different filter combinations. \par

We warn the reader that the specific values of the statistical quantities ($\sigma$, outlier percentages) quoted in this paper refer to the samples analysed here and may have variations for other galaxy samples with different redshifts and SED type distributions. Even so, as all our tests are internally consistent, our results constitute a useful reference to assess the ability to recover galaxy redshifts using different {\em JWST} filter combinations and ancillary data. \par

\section{Results}\label{sec:results}

In this section we present our results for the three galaxy samples analysed in this work. 

\subsection{Results for Sample 1}\label{subsec:analysis_zspec}

\begin{deluxetable*}{ccccc}[ht!]
\tablecaption{Statistical properties of Sample 1 for all the different {\em JWST} and ancillary data broad-band combinations. The last two columns indicate  the mean of $(z_{\mathrm{phot}}-z_{\mathrm{spec}})/(1+z_{\mathrm{spec}})$ and the r.m.s. ($\sigma$) of $|z_{\mathrm{phot}}-z_{\mathrm{spec}}|/(1+z_{\mathrm{spec}})$, respectively. The second column provides the number and percentages of $z=1-4$ sources that become outliers, i.e., objects beyond $3\sigma$ in the $z_{\mathrm{phot}}- z_{\mathrm{spec}}$ relation, with $\sigma=0.058$, which is the minimum $\sigma$ obtained with all the filter combinations considered for this galaxy sample. The total original number of $z=1-4$ galaxies in Sample 1 is 1306.  \label{tab:contamination}}
\tablecolumns{4}
\tablenum{5}
\tablewidth{0pt}
\tablehead{
\colhead{Bands} &
\colhead{$z=1-4$}&
\colhead{$<\delta z>$}&
\colhead{$\sigma_{|\delta z|}$}
}
\startdata
8 NIRCam broad bands &  134 (10.3\%)& -0.024 & 0.115\\
{\em HST} F435W + {\em HST} F606W + 8 NIRCam bands & 93 (7.1\%) & -0.032 & 0.076\\
VLT $U$ + {\em HST} F435W, F606W + 8 NIRCam & 89 (6.8\%) & -0.043 & 0.058\\
8 NIRCam bands + MIRI F560W, F770W & 116 (8.9\%) & -0.037 & 0.066\\
8 NIRCam bands + MIRI F560W & 120 (9.2\%) & -0.029 & 0.077\\
8 NIRCam bands + MIRI F770W & 128 (9.8\%) & -0.037 & 0.067\\
{\em HST} F435W, F606W + 8 NIRCam  bands + MIRI F560W, F770W & 109 (8.3\%) & -0.042 & 0.062\\
{\em HST} F435W, F606W + 8 NIRCam  bands + MIRI F560W & 93 (7.1\%) & -0.034 & 0.060\\
{\em HST} F435W, F606W + 8 NIRCam  bands + MIRI F770W & 106 (8.1\%) & -0.042 & 0.064\\
\enddata
\end{deluxetable*}

\begin{figure*}[ht!]
\center{
\includegraphics[trim={0cm 5cm 0cm 5.5cm},clip,width=1\linewidth, keepaspectratio]{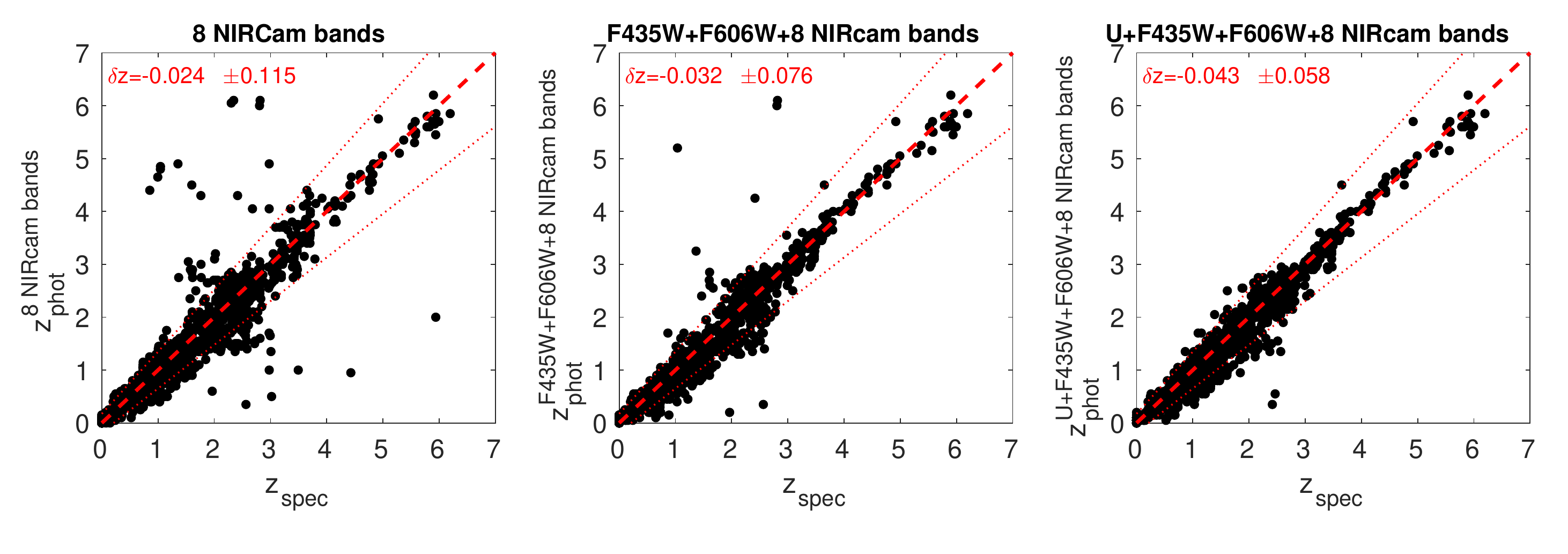}}
\caption{Photometric versus spectroscopic redshifts for galaxies in Sample 1.  \textit{Left}: photometric redshifts obtained with 8 NIRCam broad bands.  \textit{Centre}:  photometric redshifts obtained with the {\em HST} F435W, {\em HST} F606W and 8 NIRCam broad bands. \textit{Right}: the same as in the central panel, but considering also the VLT $U$ band for computing photometric redshifts.  The dotted lines in each panel delimit the $3\sigma$  limit of the $|z_{\mathrm{phot}}-z_{\mathrm{spec}}|/(1+z_{\mathrm{spec}})$ distribution, with $\sigma=$0.058, which is the minimum value of $\sigma$ that we get for Sample 1 with all the considered filter combinations. \label{fig:NIRCam_HST_U_zspec}}
\end{figure*}

\begin{figure*}[ht!]
\center{
\includegraphics[trim={0cm 5cm 0cm 6cm},clip,width=1\linewidth, keepaspectratio]{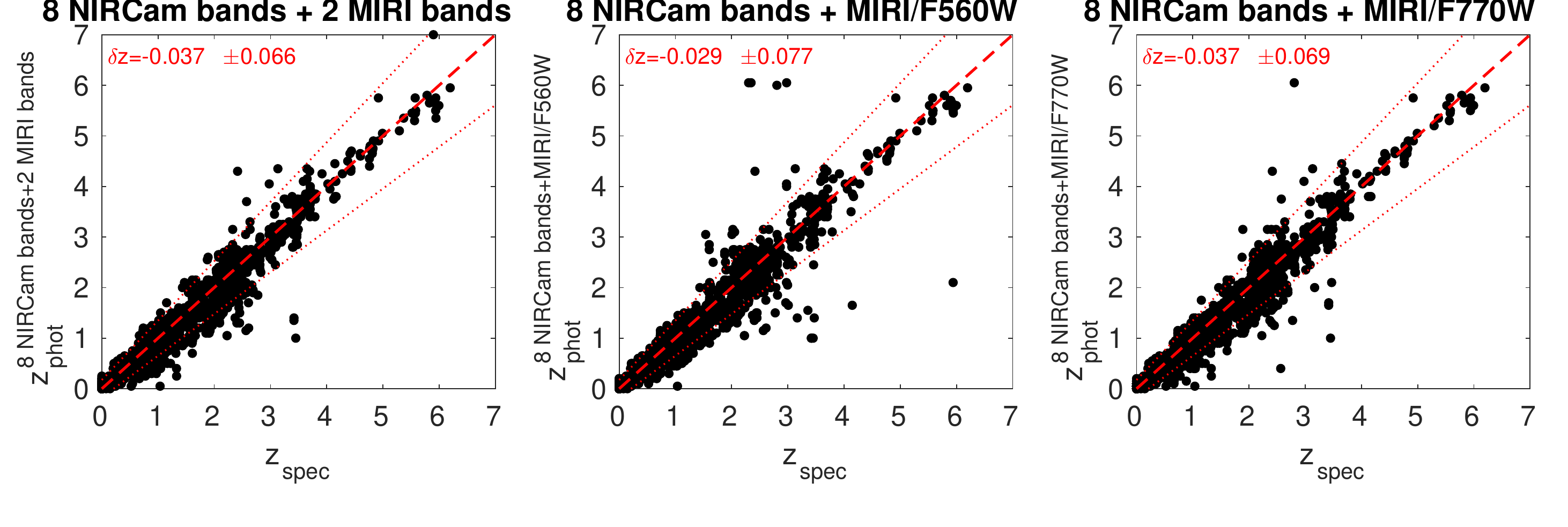}}
\caption{Photometric versus spectroscopic redshifts for galaxies in Sample 1. \textit{Left}: photometric redshifts obtained with 8 NIRCam broad bands and MIRI F560W, F770W. \textit{Centre}: Photometric redshifts obtained with 8 NIRCam broad bands and MIRI F560W only. \textit{Right}: Photometric redshift  obtained with 8 NIRCam broad bands and MIRI F770W only. The dotted lines in each panel delimit the $3\sigma$  limit of the $|z_{\mathrm{phot}}-z_{\mathrm{spec}}|/(1+z_{\mathrm{spec}})$ distribution, with $\sigma=$0.058, which is the minimum value of $\sigma$ that we get for Sample 1 with all the considered filter combinations.\label{fig:NIRCam_MIRI_zspec}}
\end{figure*}

\begin{figure*}[ht!]
\center{
\includegraphics[trim={0cm 5cm 0cm 5.5cm},clip,width=1\linewidth, keepaspectratio]{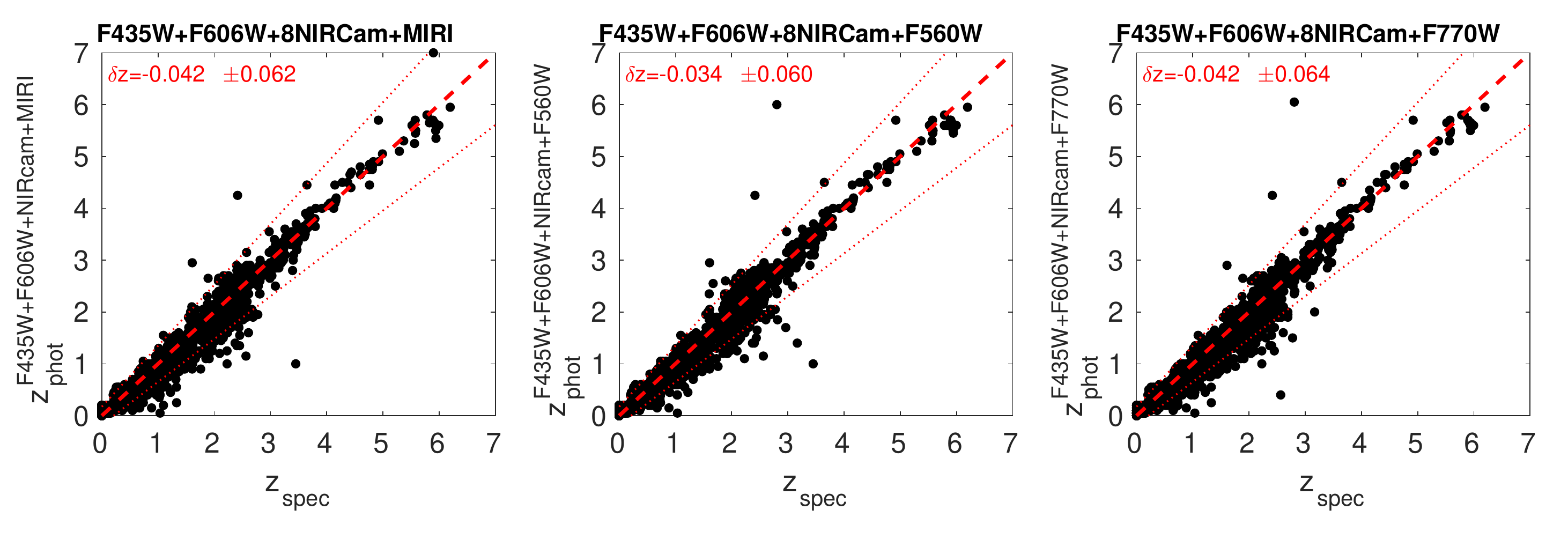}}
\caption{Photometric versus spectroscopic redshifts for galaxies in Sample 1. \textit{Left}: photometric redshifts obtained with {\em HST} F345W, {\em HST} F606W, 8 NIRCam broad bands, and MIRI F560W, F770W. \textit{Centre}: Photometric redshifts obtained with {\em HST} F345W, {\em HST} F606W, 8 NIRCam broad bands and MIRI F560W. \textit{Right}: Photometric redshifts obtained with {\em HST} F345W, F606W, 8 NIRCam broad bands and MIRI F770W. The dotted lines in each panel delimit the $3\sigma$  limit of the $|z_{\mathrm{phot}}-z_{\mathrm{spec}}|/(1+z_{\mathrm{spec}})$ distribution, with $\sigma=$0.058, which is the minimum value of $\sigma$ that we get for Sample 1 with all the considered filter combinations.\label{fig:NIRCam_HST_MIRI_zspec}}
\end{figure*}

In Figures \ref{fig:NIRCam_HST_U_zspec} to \ref{fig:NIRCam_HST_MIRI_zspec} we show the comparison between the spectroscopic and our derived photometric redshifts obtained with different filter combinations for Sample 1. For each case, we quote the mean of the $(z_{\mathrm{phot}}-z_{\mathrm{spec}})/(1+z_{\mathrm{spec}})$ distribution and the r.m.s. of  $|z_{\mathrm{phot}}-z_{\mathrm{spec}}|/(1+z_{\mathrm{spec}})$, computed considering all galaxies in the sample.  We identified 3$\sigma$ outliers using as reference $\sigma=$0.058, which is the minimum of all the $\sigma$ values that we get for Sample 1. This galaxy sample is particularly useful to test the percentage of low-$z$ galaxies ($z<4$) that leak to wrong redshifts due to failures in the redshift recovery. The percentages of these low-$z$ sources that are leaking outliers for each filter combination is listed in Table~\ref{tab:contamination}. For this analysis, we considered only the redshift bin $z=1-4$, avoiding lower $z$ galaxies, as the MIRI filters are not very suitable to trace the direct stellar emission at such low $z$. Note that, although these obtained outlier percentages apply particularly to this galaxy sample, the changes obtained with different passbands are of general validity and show the impact of incorporating different filters on the redshift recovery.   \par

Firstly, we analysed the case of using the 8 NIRCam broad bands with and without ancillary {\em HST} and ground-based $U$-band data (Figure~\ref{fig:NIRCam_HST_U_zspec}). We found that, when using only the 8 NIRCam broad bands, the percentage of $z=1-4$ sources that are $3\sigma$ outliers is of 10.3\%, including several extreme cases which appear as $z>4$ contaminants.  Besides, the r.m.s. of the $|z_{\mathrm{phot}}-z_{\mathrm{spec}}|/(1+z_{\mathrm{spec}})$ distribution is high ($\sigma=0.115$) compared with the typical precision that can currently be achieved for low-$z$ galaxies in deep surveys. When adding shorter-wavelength {\em HST} data, the normalised redshift difference distribution r.m.s. is significanlty lower, the outlier percentage decreases to 7.1\%, and the majority of the most extreme redshift outliers are corrected. When adding also the $U$ band, the r.m.s. reaches a minimum ($\sigma=0.058$) and so does the outlier percentage (6.8\%), leaving basically no catastrophic outlier cases.  This is because it is necessary to cover wavelengths shorter than $0.7 \, \rm \mu m$ in order to distinguish between the Lyman and $4000 \, \rm \AA$ breaks at $z<6$. The improvement on the photometric redshift estimation is also due to the mere increase in the wavelength coverage given by the larger number of bands. Besides, the incorporation of the {\em HST} and $U$-band data to the NIRCam photometry reduces the photometric redshift dispersion ($\sigma$) to about a half (Fig.~\ref{fig:NIRCam_HST_U_zspec}). \par

Secondly, we tested the photometric redshift estimation when complementing the NIRCam photometry with MIRI photometry, instead of shorter wavelength data (Figure \ref{fig:NIRCam_MIRI_zspec}). When we incorporate both MIRI F560W and F770W bands, the resulting normalised redshift difference distribution r.m.s. is significantly lower ($\sigma=0.066$) than in the case with NIRCam data alone, indicating that the presence of MIRI photometry can mitigate the absence of short-wavelength data. (Table~\ref{tab:contamination}).\par

Lastly, we considered the situation in which {\em HST} ancillary data along with NIRCam and MIRI data are available (Figure \ref{fig:NIRCam_HST_MIRI_zspec}). We explicitly tested a combination without $U$-band photometry, as ground-based $U$-band observations with matching depth to  {\em JWST} observations will be time-consuming and difficult to achieve. The results are rather similar to the case with NIRCam and MIRI bands considered together: the obtained r.m.s. is $\sigma=0.062$, with the fraction of redshift outliers being slightly reduced. Interestingly, when only NIRCam and MIRI photometry are considered, the F770W data improves the overall redshift statistics.  Instead, when {\em HST} data are also available, the incorporation of F560W alone provides slightly better results. (This is related to how the $\chi^2$-fitting procedure works, typically giving more weight to longer wavelength data where the sources are brighter).\par

To conclude, the 8 NIRCam broad bands alone are not adequate to obtain sufficiently good photometric redshifts at $z<4$. Although for most galaxies the derived redshifts are correct, the overall obtained statistics is poor compared with the typical $z_{\rm phot}$ quality that can currently be achieved in deep surveys at these redshifts.  Observations at wavelengths shorter than $0.7 \, \rm \mu m$ are necessary to remove the most extreme redshift outliers. Incorporating MIRI longer-wavelength photometry instead can mitigate the absence of short-wavelength observations. Having a complete optical and infrared coverage up to $7.7 \, \rm \mu m$  will allow us to keep the $z<4$ redshift statistics well under control.

\begin{deluxetable*}{ccccc}[ht!]
\tablecaption{Statistical properties for Sample 2 for all the different {\em JWST} and ancillary data broad-band combinations. The last two columns indicate  the mean of $(z_{\mathrm{phot}}-z_{\mathrm{spec}})/(1+z_{\mathrm{spec}})$ and the r.m.s. ($\sigma$) of $|z_{\mathrm{phot}}-z_{\mathrm{spec}}|/(1+z_{\mathrm{spec}})$, respectively. Columns 2 and 3 provide the numbers and percentages of galaxies leaking towards lower redshifts (see Fig. \ref{fig:NIRCam_HST_U_zspec}). These outliers are defined as galaxies beyond  3$\sigma$, with $\sigma=$0.065, which is the minimum $\sigma$ obtained with the different filter combinations considered for this galaxy sample. The total original number of $z=4-5$ ($z=5-7$) galaxies in Sample 2 is 887 (488). \label{tab:outliers}}
\tablecolumns{5}
\tablenum{6}
\tablewidth{0pt}
\tablehead{
\colhead{Bands} &
\colhead{$z=4-5$}&
\colhead{$z=5-7$}&
\colhead{$<\delta z>$} &
\colhead{$\sigma_{<|\delta z|>}$}
}

\startdata
8 NIRCam broad bands & 190 (21.4$\%$) & 46 (9.4$\%$) & -0.111 & 0.218\\
{\em HST} F435W, F606W + 8 NIRCam bands & 21 (2.4$\%$) & 36 (8.0$\%$) & -0.023 & 0.119\\
VLT $U$ + {\em HST} F435W, F606W+ 8 NIRCam  bands & 23 (2.6$\%$)  & 37 (8.2$\%$)  & -0.024 & 0.122\\
8 NIRCam  bands + MIRI F560W, F770W & 68 (7.7$\%$)& 14 (3.1$\%$) & -0.034 & 0.099\\
8 NIRCam  bands + MIRI F560W & 120 (13.5$\%$) & 41 (9.1$\%$) & -0.072 & 0.169\\
8 NIRCam  bands + MIRI F770W & 90 (10.1$\%$) & 31 (6.9$\%$) & -0.049 & 0.127\\
{\em HST} F435W, F606W + 8 NIRCam  bands + MIRI F560W, F770W & 11 (1.2$\%$) & 7 (1.4$\%$)  & -0.004 & 0.065\\
{\em HST} F435W, F606W + 8 NIRCam  bands + MIRI F560W & 20 (2.2$\%$) & 32 (6.6$\%$)  & -0.019 & 0.113\\
{\em HST} F435W, F606W + 8 NIRCam  bands + MIRI F770W & 17 (1.9$\%$) & 18 (4.0$\%$)  & -0.011 & 0.083\\
\enddata
\end{deluxetable*}

\begin{figure*}[ht!]
\center{
\includegraphics[trim={0cm 5cm 0cm 5cm},clip,width=1\linewidth, keepaspectratio]{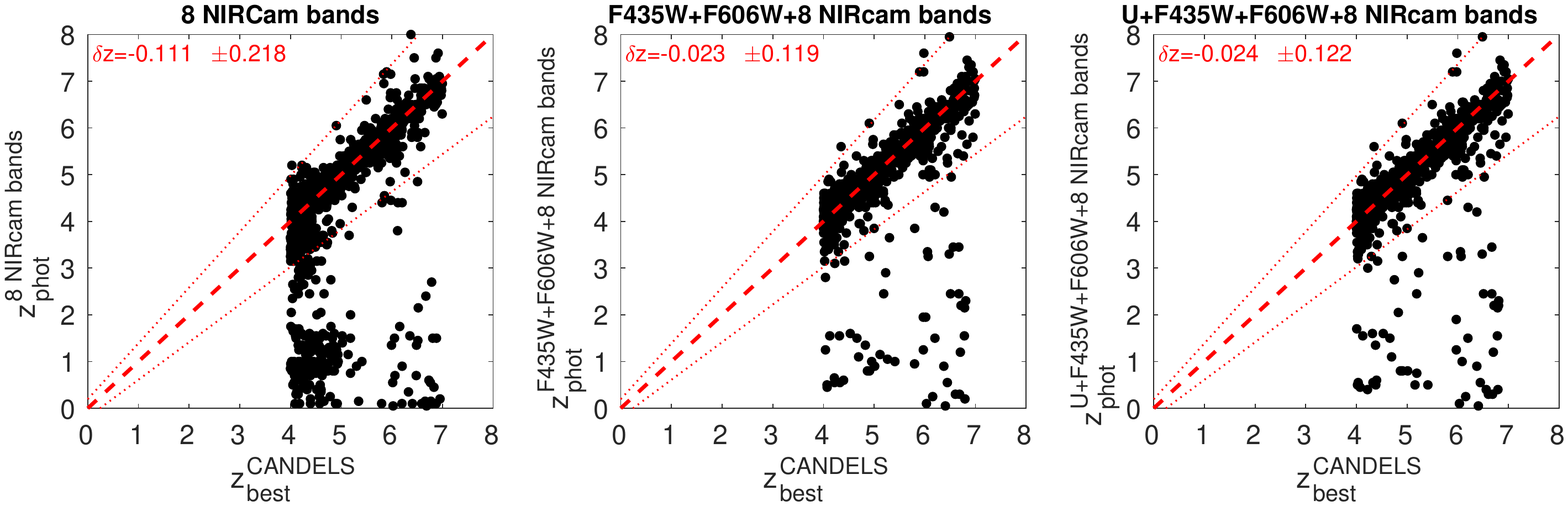}}
\caption{Our derived photometric redshifts versus CANDELS consensus photometric redshifts for galaxies in Sample 2. \textit{Left}: Photometric redshifts obtained with 8 NIRCam broad bands. \textit{Centre}: Photometric redshifts obtained with {\em HST} F435W, F606W and 8 NIRCam broad bands. \textit{Right}: Photometric redshifts obtained with the same bands as in the central panel plus the VLT $U$ band. The dotted lines in each panel delimit the $3\sigma$  limit of the $|z_{\mathrm{phot}}-z_{\mathrm{fiduc.}}|/(1+z_{\mathrm{fiduc.}})$ distribution, with $\sigma=$0.065, which is the minimum value of $\sigma$ that we get for Sample 2 with all the considered filter combinations.\label{fig:NIRCam_HST_U}}
\end{figure*}

\subsection{Results for Sample 2}\label{subsec:analysis_zphot}

\begin{figure*}[ht!]
\center{
\includegraphics[trim={0cm 5cm 0cm 5cm},clip,width=1\linewidth, keepaspectratio]{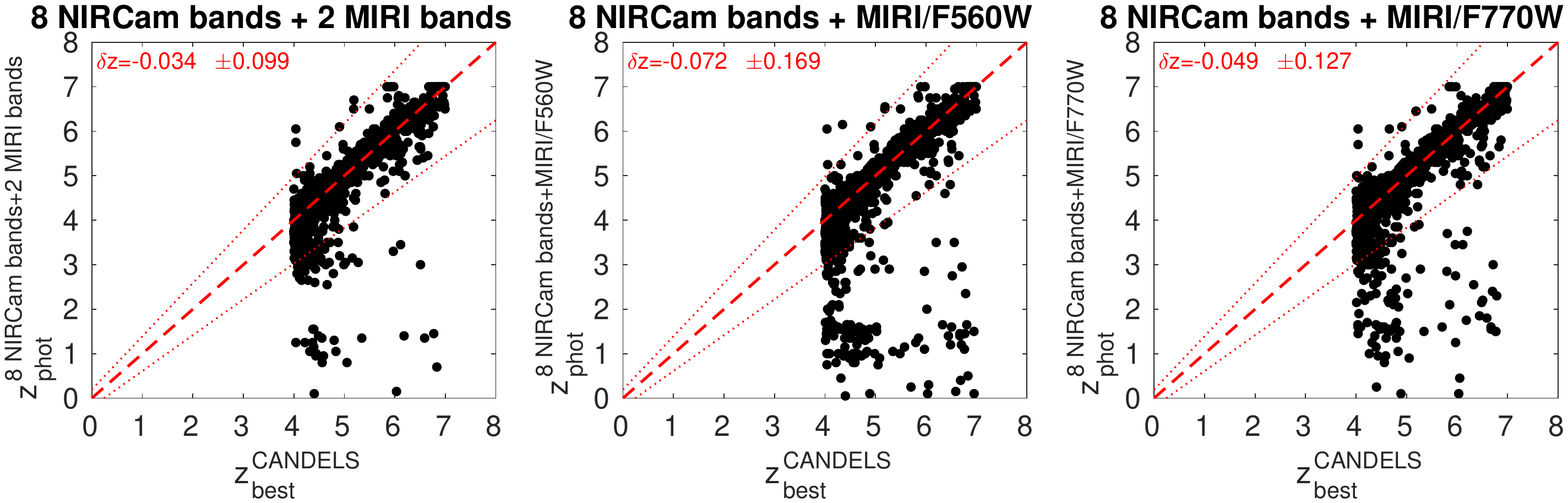}}
\caption{Our derived photometric redshifts versus CANDELS consensus photometric redshifts for galaxies in Sample 2. \textit{Left}: photometric redshifts obtained with 8 NIRCam broad bands and MIRI F560W, F770W. \textit{Centre}: Photometric redshifts obtained with 8 NIRCam broad bands and MIRI F560W only. \textit{Right}: Photometric redshifts obtained with 8 NIRCam broad bands and MIRI F770W only. The dotted lines in each panel delimit the $3\sigma$  limit of the $|z_{\mathrm{phot}}-z_{\mathrm{fiduc.}}|/(1+z_{\mathrm{fiduc.}})$  distribution, with $\sigma=$0.065, which is the minimum value of $\sigma$ that we get for Sample 2 with all the considered filter combinations.\label{fig:NIRCam_MIRI}}
\end{figure*}

\begin{figure*}[ht!]
\center{
\includegraphics[trim={0cm 5cm 0cm 5.5cm},clip,width=1\linewidth, keepaspectratio]{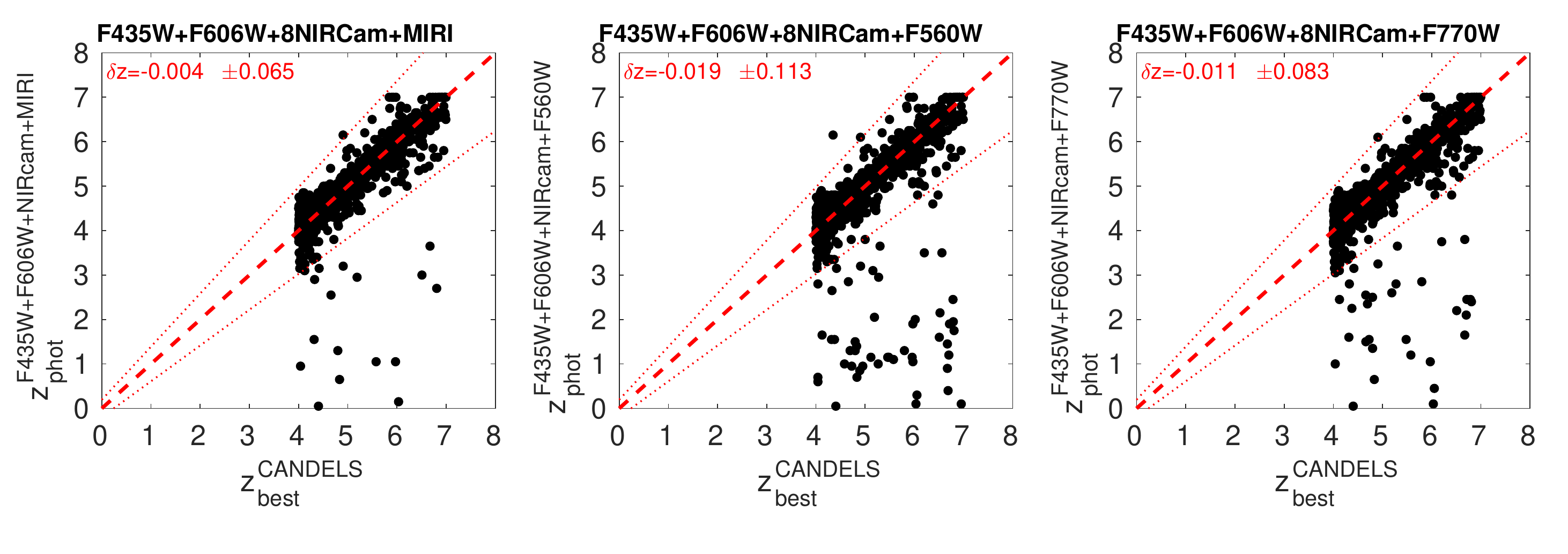}}
\caption{Our derived photometric redshifts versus CANDELS consensus photometric redshifts for galaxies in Sample 2. \textit{Left}: photometric redshifts obtained with {\em HST} F345W, F606W, 8 NIRCam broad bands, and MIRI F560W, F770W. \textit{Centre}: Photometric redshifts obtained with {\em HST} F345W,F606W, 8 NIRCam broad bands, and MIRI F560W. \textit{Right}: Photometric redshift is obtained with {\em HST} F345W, F606W, 8 NIRCam broad bands and MIRI F770W. The dotted lines in each panel delimit the $3\sigma$  limit of the $|z_{\mathrm{phot}}-z_{\mathrm{fiduc.}}|/(1+z_{\mathrm{fiduc.}})$  distribution, with $\sigma=$0.065, which is the minimum value of $\sigma$ that we get for Sample 2 with all the considered filter combinations.\label{fig:NIRCam_HST_MIRI}}
\end{figure*}

In Figures \ref{fig:NIRCam_HST_U} to \ref{fig:NIRCam_HST_MIRI} we show the comparison between our recovered photometric redshifts and fiducial CANDELS consensus redshifts for galaxies in Sample 2 (all of which have fiducial redshifts $z>4$).  As we did before, we derived $|z_{\mathrm{phot}}-z_{\mathrm{fiduc.}}|/(1+z_{\mathrm{fiduc.}})$ for all our galaxies and we identified 3$\sigma$ outliers, where we adopted $\sigma=0.065$, which is the smallest r.m.s. value that we obtained with the different filter combinations. The percentages of outliers in each case are listed in Table \ref{tab:outliers}.  \par

\begin{deluxetable*}{crrrr}[ht!]
\tablecaption{Number of outliers among our 864 Sample 3 galaxies simulated with BC03 models and manual addition of emission lines for different {\em JWST} filter combinations and assumed $S/N$ values at F150W. Outliers are defined as galaxies with $|z_{\mathrm{phot}}-z_{\mathrm{fiduc.}}|/(1+z_{\mathrm{fiduc.}})>0.15$.  \label{tab:outliers_z7}}
\tablecolumns{5}
\tablewidth{0pt}
\tablehead{
\colhead{Bands} &
\colhead{$S/N=10$} &
\colhead{$S/N=5$} &
\colhead{$S/N=5$} &
\colhead{$S/N=3$} \\
\colhead{} &
\colhead{} &
\colhead{} &
\colhead{(only EL)\tablenotemark{a}} &
\colhead{}
}
\startdata
8 NIRCam broad bands & 1 ($0.1\%$) & 24 ($2.8\%$) &  20\tablenotemark{b} (3.5\%)& 87 (10.1\%) \\
8 NIRCam bands + MIRI F560W, F770W & 1 ($0.1\%$) & 5 ($0.6\%$) & 3\tablenotemark{b} (0.5\%) & 26 (3.0\%)\\
8 NIRCam bands + MIRI F560W & 1 ($0.1\%$) & 7 ($0.8\%$)& 3\tablenotemark{b} (0.5\%)& 29 (3.4\%)\\
8 NIRCam bands + MIRI F770W & 2 ($0.2\%$) & 16 ($1.9\%$)  & 14\tablenotemark{b} (2.4\%) & 45 (5.2\%)\\
\enddata
\tablenotetext{a}{only templates with emission lines (EL).}
\tablenotetext{b}{out of a total of 576 galaxies with emission lines. The percentages refer to this denominator.}
\end{deluxetable*}

\begin{deluxetable*}{crrrr}[ht!]
\tablecaption{Number of outliers among our 1260 Sample 3 galaxies simulated with {\em Yggdrasil} models for different {\em JWST} filter combinations and assumed $S/N$ values at F150W. Outliers are defined as galaxies with $|z_{\mathrm{phot}}-z_{\mathrm{fiduc.}}|/(1+z_{\mathrm{fiduc.}})>0.15$.  \label{tab:outliers_z7_Ygg}}
\tablecolumns{5}
\tablewidth{0pt}
\tablehead{
\colhead{Bands} &
\colhead{$S/N=10$} &
\colhead{$S/N=5$} &
\colhead{$S/N=5$} &
\colhead{$S/N=3$} \\
\colhead{} &
\colhead{} &
\colhead{} &
\colhead{(only EL)\tablenotemark{a}} &
\colhead{}
}
\startdata
8 NIRCam bands & 11 (0.9\%) & 257 (20.4\%) & 118\tablenotemark{b}  (49.2\%)  & 560 (44.4\%) \\
8 NIRCam bands + MIRI F560W, F770W & 1 (0.1\%)  & 152 (12.1\%) & 41\tablenotemark{b} (17.1\%) & 462 (36.7\%) \\
8 NIRCam bands + MIRI F560W & 13 (1.0\%) & 188 (14.9\%) & 68\tablenotemark{b} (28.3\%) & 490 (38.9\%) \\
8 NIRCam bands + MIRI F770W & 8 (0.6\%) & 196 (15.6\%) & 79\tablenotemark{b} (32.9\%) & 508 (40.3\%) \\
\enddata
\tablenotetext{a}{only templates with emission lines (EL).}
\tablenotetext{b}{out of a total of 240 galaxies with emission lines. The percentages refer to this denominator.}
\end{deluxetable*}

As for Sample 1, our first step consisted of analysing the resulting photometric redshifts obtained using the 8 NIRCam broad bands with and without the {\em HST} and $U$-band data (Figure \ref{fig:NIRCam_HST_U}). We found that when using only the 8 NIRCam broad bands the percentage of outliers for Sample 2 is high. We did not show the percentage of low-$z$ contaminants because of the sample selection, which by construction only has $z>4$ objects. The most evident effect in Sample 2 is the leakage of high-$z$ sources to low-$z$ values produced by the insufficient photometric coverage. This is particularly important for galaxies with input redshifts $z=4-5$: a bit more than 20\% of them are found $>3\sigma$ away at lower redshifts (Table~\ref{tab:outliers}).\par

The redshift estimation is better at higher redshifts ($z=5-7$). The percentage of sources leaking to lower $z$ is of only $\sim 9\%$. As we discussed before, the Lyman break enters  the shortest-wavelength NIRCam band at $z\sim5-6$, and therefore, the degeneracy between the $4000 \, \rm \AA$ and Lyman breaks occurs mainly at lower redshifts. For this reason, adding the {\em HST} F435W and 606W photometry does not have an important effect on the redshift estimation at these higher redshifts, but it does at $z=4-5$, where it reduces the percentage of leaking sources to only $\sim 2\%$.  The overall redshift distribution dispersion reduces to a half when the {\em HST} data is considered along the NIRCam photometry (Fig.~\ref{fig:NIRCam_HST_U}).\par

The addition of the $U$ band does not produce any obvious effect in the redshift estimation for Sample 2, once the {\em HST} photometry is also considered. The role of the $U$ band is more important for identifying low-redshift contaminants rather than preventing the leakage of high-$z$ sources to low $z$, because galaxies at z$>$4 are very faint or not detected in this band. As we will discuss later, the {\em HST} optical bands have a similar effect at z$>$5. \par

As a second step, we obtained photometric redshifts with the 8 NIRCam broad bands and two MIRI bands F560W and F770W, considered together and one at a time  (Figure \ref{fig:NIRCam_MIRI}). Adding both MIRI bands significantly improves the photometric redshift estimates by decreasing the percentage of leaking sources to $<8\%$ at $z=4-5$ and $<4\%$ at $z=5-7$. Among the remaining $z=4-5$ leaking sources, some are placed at $z\sim1$ and a few at higher redshifts, i.e., $z\sim6.5$. When considering one MIRI band at a time, F770W appears to be more effective at improving the photometric redshift estimation than F560W, but it is the combined effect of both MIRI bands which produces the substantial reduction in the fraction of leaking galaxies.\par

\begin{figure*}[ht!]
\center{
\includegraphics[width=1\linewidth, keepaspectratio]{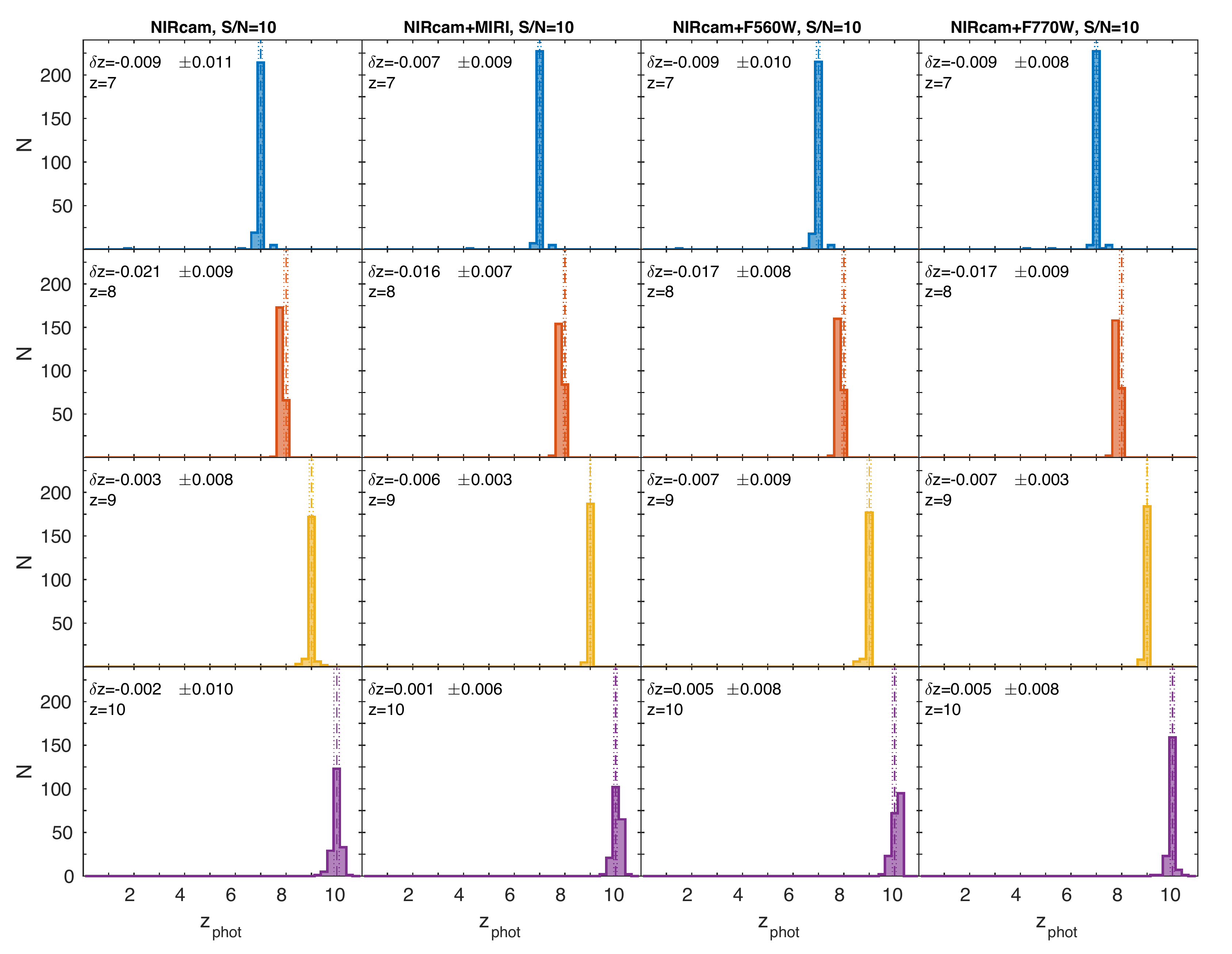}}
\caption{Photometric redshifts obtained for the  BC03 simulated galaxies with (F150W) $S/N=10$ at different fixed redshifts. \textit{From top to bottom}: redshifts $z=7, 8, 9$ and 10.  Photometric redshifts in each column are obtained with different combinations of bands. \textit{From left to right:} 8 NIRCam broad bands; 8 NIRCam broad bands, MIRI F560W and MIRI F770W; 8 NIRCam broad bands and MIRI F560W only; 8 NIRCam broad bands and MIRI F770W only. Each row corresponds to one of the four specific input redshifts. The vertical lines indicate the $3\sigma$ interval around the mean normalised redshift difference.  \label{fig:z7_StN10}}
\end{figure*}

\begin{figure*}[ht!]
\center{
\includegraphics[width=1\linewidth, keepaspectratio]{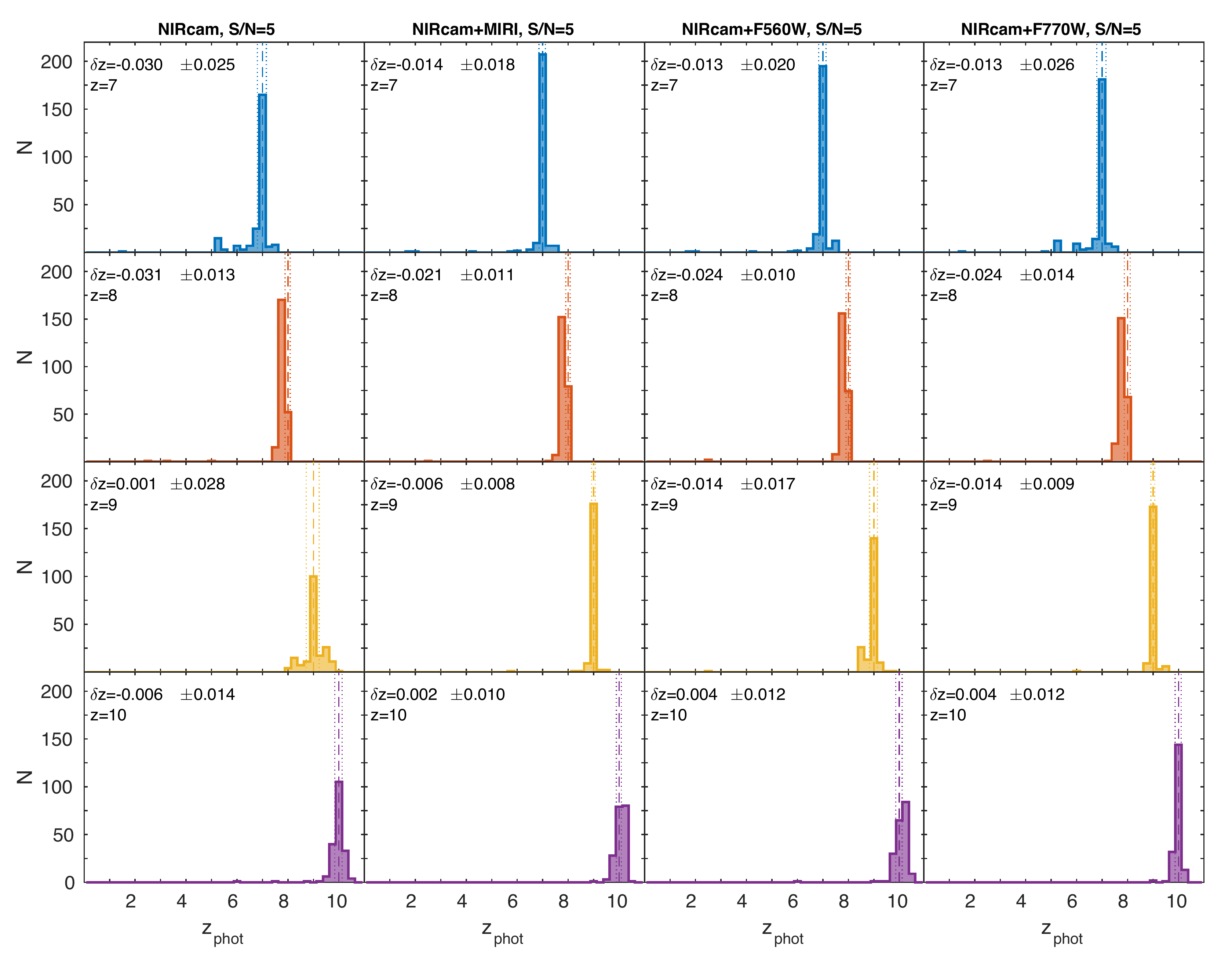}}
\caption{The same as Fig. \ref{fig:z7_StN10}, but for (F150W)  $S/N=5$. \label{fig:z7_StN5}}
\end{figure*}

\begin{figure*}[ht!]
\center{
\includegraphics[width=1\linewidth, keepaspectratio]{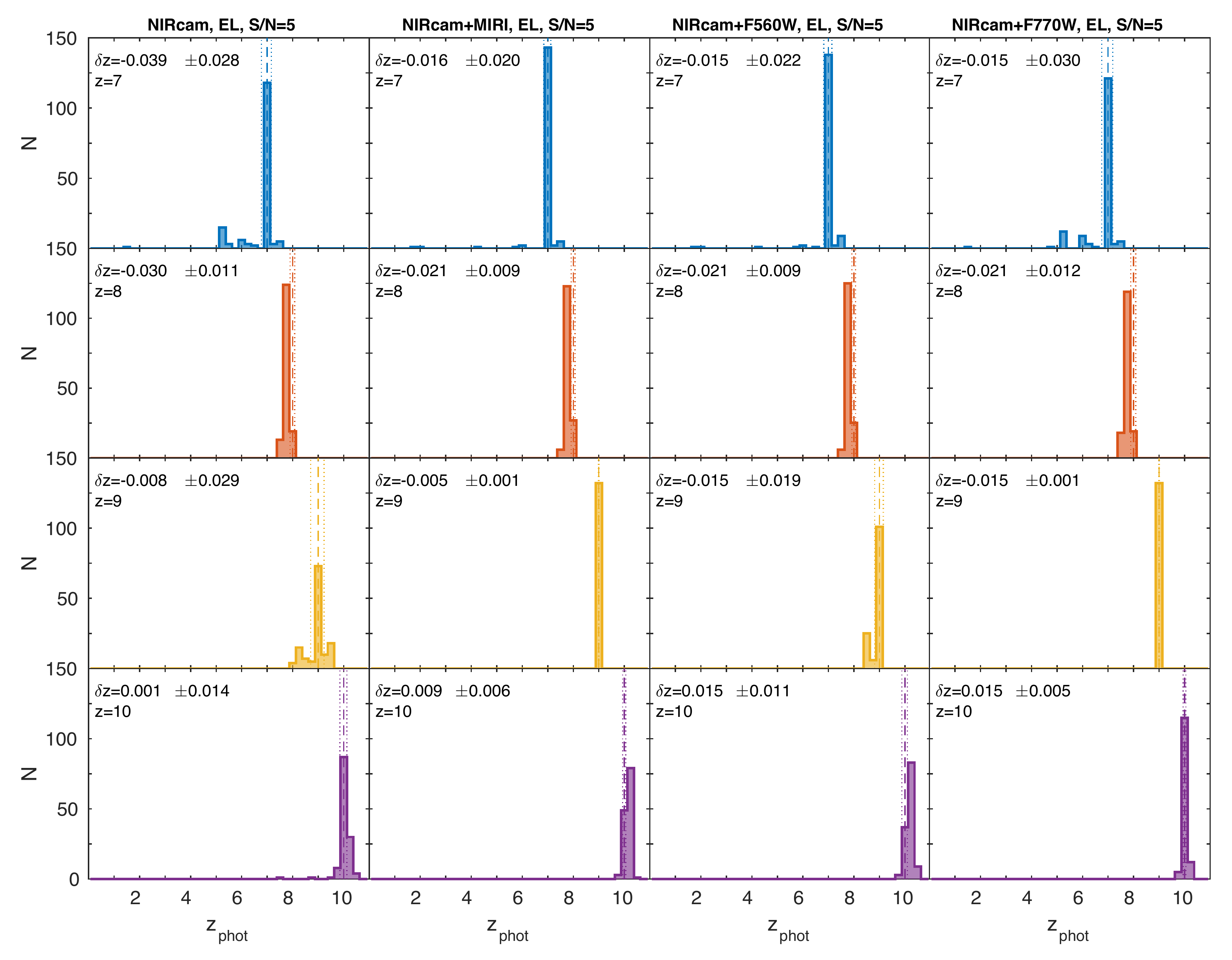}}
\caption{The same as Fig. \ref{fig:z7_StN5}, but only for galaxies with emission lines. \label{fig:z7_StN5el}}
\end{figure*}

\begin{figure*}[ht!]
\center{
\includegraphics[width=1\linewidth, keepaspectratio]{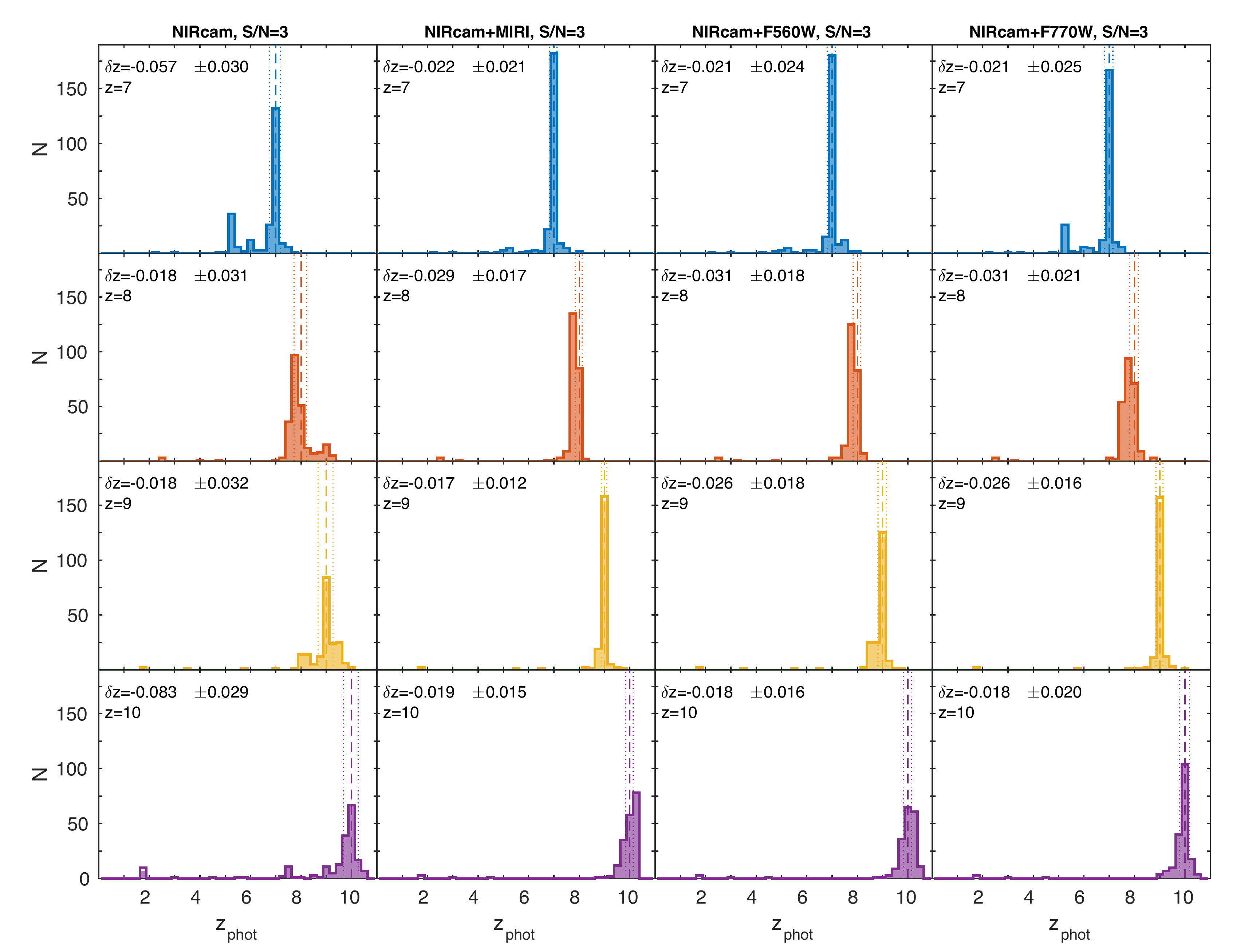}}
\caption{The same as Fig. \ref{fig:z7_StN10}, but for (F150W)  $S/N=3$. \label{fig:z7_StN3}}
\end{figure*}

Adding MIRI photometry appears to be more important than adding {\em HST} photometry to minimise the dispersion in the obtained redshift distribution: the resulting r.m.s. in the case of NIRCam and MIRI data is $\sigma=0.099$, which is slightly lower than for the {\em HST} and NIRCam filter combination ($\sigma=0.119$). \par

Lastly, we considered {\em HST}, NIRCam and MIRI photometry altogether to obtain photometric redshifts (Figure \ref{fig:NIRCam_HST_MIRI}). As expected, using all these bands jointly provides the best results in the identification of z$>4$ galaxies.  The presence of {\em HST} photometry at $\lambda<0.6-0.7 \, \rm \mu m$  helps to correctly identify the Lyman break at $z=4-5$, while the MIRI mid-infrared bands provide a better wavelength coverage to correctly identify sources at z$=5-7$. The resulting redshift dispersion gets a minimum value of $\sigma=0.065$ and the fraction of $z=4-5$ sources leaking beyond $3\sigma$ in redshift decreases to $\sim 1\%$.\par 

As before, the MIRI F770W filter appears to have a more important role in improving the redshift estimation than MIRI F560W, but both bands are necessary to properly identify all galaxies at $z=4-7$. To sum up, the short-wavelength {\em HST} data and long-wavelength MIRI data have complementary roles in the correct identification of $z>4$ galaxies. \par

We remind the reader that the photometry in our Samples 1 and 2 is of good quality ($S/N>5$ in F150W for the vast majority of galaxies; see Section~\ref{sec-jwstphotmeas}).  The redshift outlier percentages that we quote here correspond to this high-quality photometry, so they do not depend on the data quality, but are rather produced by the intrinsic limitations of the SED fitting with a limited number of photometric bands. We checked the $S/N$ of the photometry for the $z_{\rm phot}$ outliers and found no significant difference with respect to the photometry of all other sources, confirming that Samples 1 and 2 results are not affected by the photometric $S/N$. \par

\subsection{Results for Sample 3}\label{subsec:analysis_highz}

With Sample 3 we aim to test the ability to recover photometric redshifts for simulated galaxies at $z=7-10$, using the NIRCam broad bands alone and in conjunction with the MIRI F560W and F770W bands. For this sample, we did not consider the incorporation of the {\em HST} short-wavelength and $U$ filters,  because these bands map rest wavelengths far blueward of the Lyman break at these redshifts, so they basically correspond to non-detections in all cases. \par

\subsubsection{Galaxies simulated with BC03 templates at $z=7-10$}
\label{sec:z7bc03}

Figures \ref{fig:z7_StN10} to \ref{fig:z7_StN3} show the distributions of the recovered photometric redshifts for each input redshift ($z=7, 8, 9$ and 10) for the Sample 3 galaxies simulated with the BC03 templates, obtained by running \textit{LePhare} with different {\em JWST} filter combinations. As in this case the input redshifts are four fixed values, it is clearer to show the output $z_{\rm phot}$ distributions for each input redshift rather than $z_{\mathrm{phot}}-z_{\mathrm{phot}}$ plots, which would contain four vertical columns each. \par

As before, in each plot we quote the mean value of the normalised redshift difference distribution $\delta z=(z_{\mathrm{phot}}-z_{\mathrm{fiduc.}})/(1+z_{\mathrm{fiduc.}})$, taking into account all galaxies. However, for Sample 3 galaxies, we adapted our method for the r.m.s. calculation. As it can be seen in Figures \ref{fig:z7_StN10} to \ref{fig:z7_StN3}, the Sample 3 redshift histograms are characterised by quite a narrow distribution around the main peak and the presence of some catastrophic failures producing secondary peaks at much lower redshifts. These catastrophic redshift failures would significantly bias the computed $\sigma$ of  $|z_{\mathrm{phot}}-z_{\mathrm{fiduc.}}|/(1+z_{\mathrm{fiduc.}})$ if they were considered in the statistics. So, for Sample 3 galaxies (both based on the BC03 and \textit{Yggdrasil} models) we explicitly excluded all galaxies with $|z_{\mathrm{phot}}-z_{\mathrm{fiduc.}}|/(1+z_{\mathrm{fiduc.}})>0.15$ for the $\sigma$ computation, and we considered those excluded galaxies to be the outliers of our sample. We provide the statistics on the outlier fractions for the BC03-SED galaxies in Table~\ref{tab:outliers_z7}. \par

For $S/N=10$, the 8 NIRCam broad bands alone allowed us to derive accurate redshift estimates, with mean values for the redshift distributions close to 0 and $\sigma \leq 0.011$ at all redshifts (Fig.~\ref{fig:z7_StN10}). The percentage of outliers quoted in  Table~\ref{tab:outliers_z7} is very small ($0.1\%$).  Adding MIRI photometry to the NIRCam photometry only slightly enhances the already excellent quality of the output redshifts, resulting in $\sigma<0.01$ at all redshifts.  These results indicate that the incorporation of the MIRI photometry to the NIRCam photometry has in practice little effect to derive photometric redshifts for $z=7-10$ galaxies with BC03-like SEDs when the NIRCam $S/N$ is very high.  \par

\begin{figure*}[ht!]
\center{
\includegraphics[width=1\linewidth, keepaspectratio]{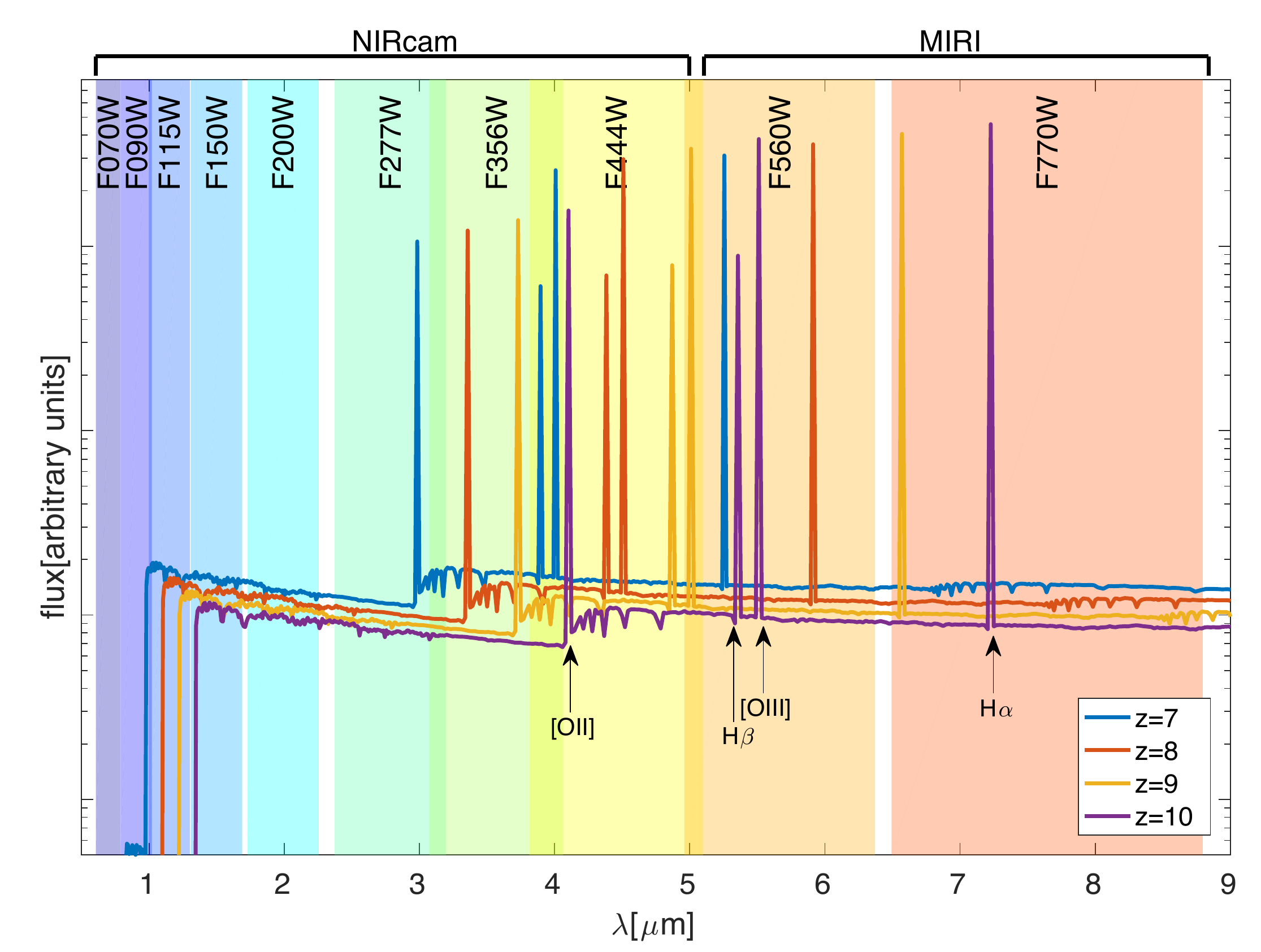}}
\caption{Four examples of SED templates from BC03 with manual emission line addition, considered for simulating galaxies for Sample 3.  The coloured vertical strips indicate the wavelength coverage of the 8 NIRCam broad bands and MIRI F560W and F770W. This Figure shows in which {\em JWST} filters the main emission lines and most important SED features are observed at high redshifts.\label{fig:SED&bands}}
\end{figure*}

The MIRI effect to the photometric redshift estimation is more evident at $S/N=5$ (Fig. \ref{fig:z7_StN5}). The NIRCam data alone are sufficient for a reasonably good redshift recovery ($\sigma \approx 0.01-0.03$), but the addition of MIRI data improves the photometric redshift quality, taking the  outlier percentage to $<1\%$ (when  F560W photometry is included). Inspection of Fig.~\ref{fig:sed_z7} suggests that this is a k-correction effect: galaxies are brighter in the MIRI bands either because they are old or because of the presence of emission lines. Indeed, both effects are important. In Fig.~\ref{fig:z7_StN5el} we show separately our results only for galaxies with emission lines: it can be seen that the effect of incorporating the MIRI photometry is similar as in the general case with $S/N=5$. Note that, on the one hand, the presence of emission lines  helps in the redshift identification, but on the other hand having prominent emission lines in contiguous broad bands makes the observed photometry mimic a less featured SED, resulting in a more difficult redshift recovery. \par

\begin{figure*}[ht!]
\center{
\includegraphics[width=1\linewidth, keepaspectratio]{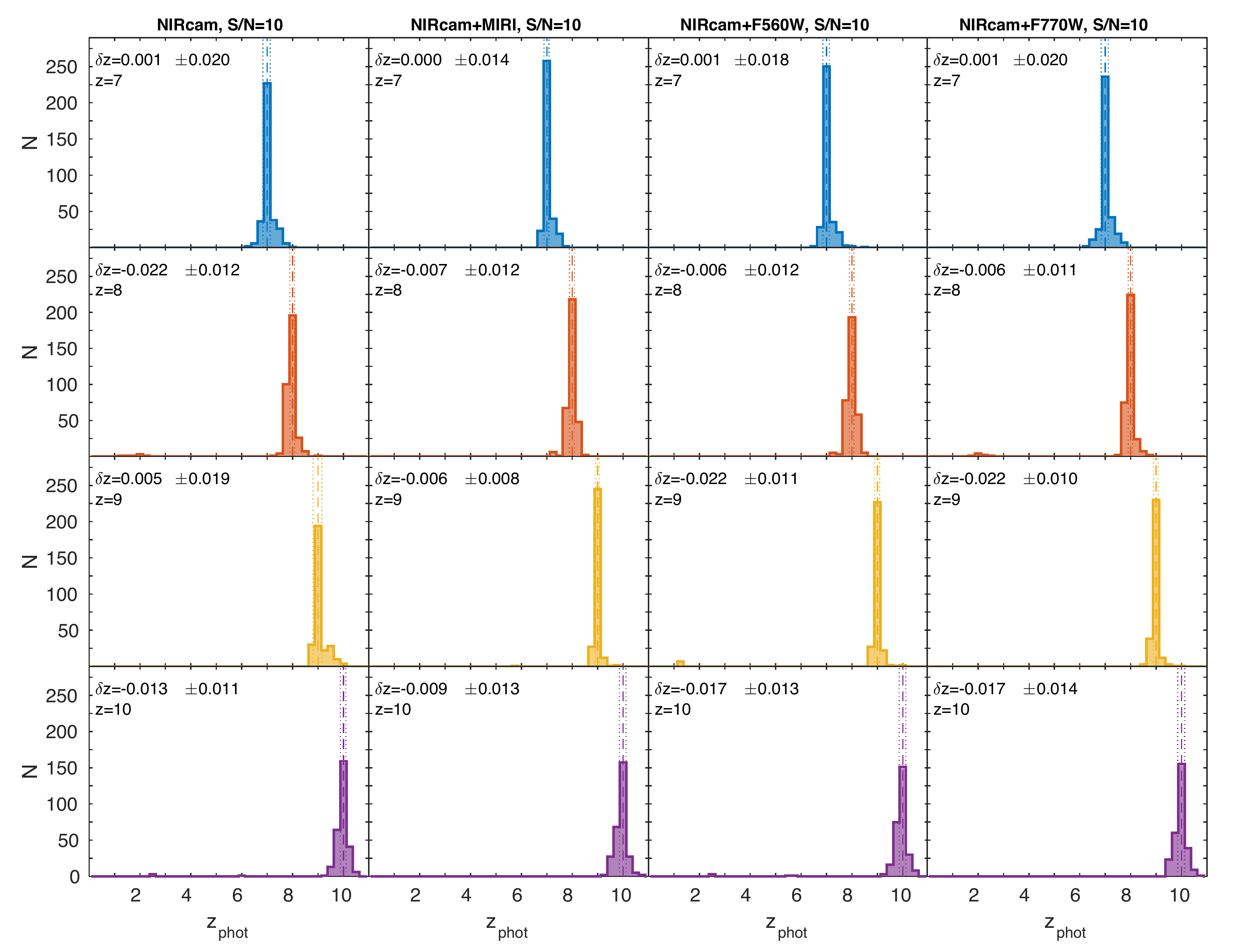}}
\caption{Photometric redshifts for the sample of {\em Yggdrasil} simulated galaxies with $S/N=10$. \textit{From top to bottom}: redshifts $z=7, 8, 9$ and 10.  Photometric redshifts in each column are obtained with different band combinations. \textit{From left to right}: 8 NIRCam broad bands; 8 NIRCam broad bands and MIRI F560W, F770W; 8 NIRCam broad bands and MIRI F560W only; 8 NIRCam broad bands and MIRI F770W only. Each row corresponds to one of the four specific input redshifts. The vertical lines indicate the $3\sigma$ interval around the mean normalised redshift difference.  \label{fig:z7_YggSN10}}
\end{figure*}

\begin{figure*}[ht!]
\center{
\includegraphics[width=1\linewidth, keepaspectratio]{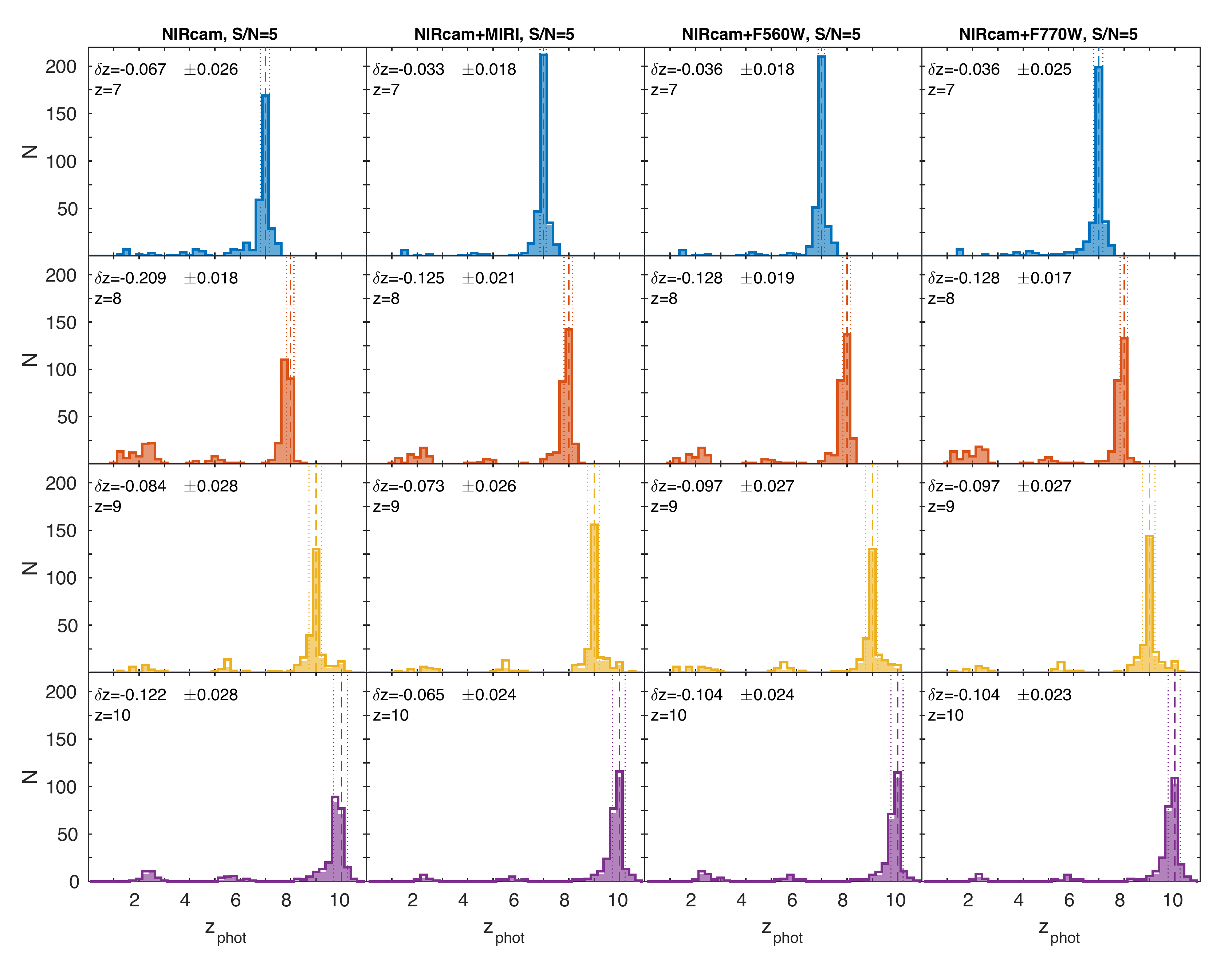}}
\caption{The same as Fig. \ref{fig:z7_YggSN10}, but for $S/N=5$. The shaded area within each histogram corresponds to objects with more than 4 detections in total in the NIRCam and MIRI bands, while empty areas correspond to sources with less than 4-band detections. \label{fig:z7_YggSN5}}
\end{figure*}

\begin{figure*}[ht!]
\center{
\includegraphics[width=1\linewidth, keepaspectratio]{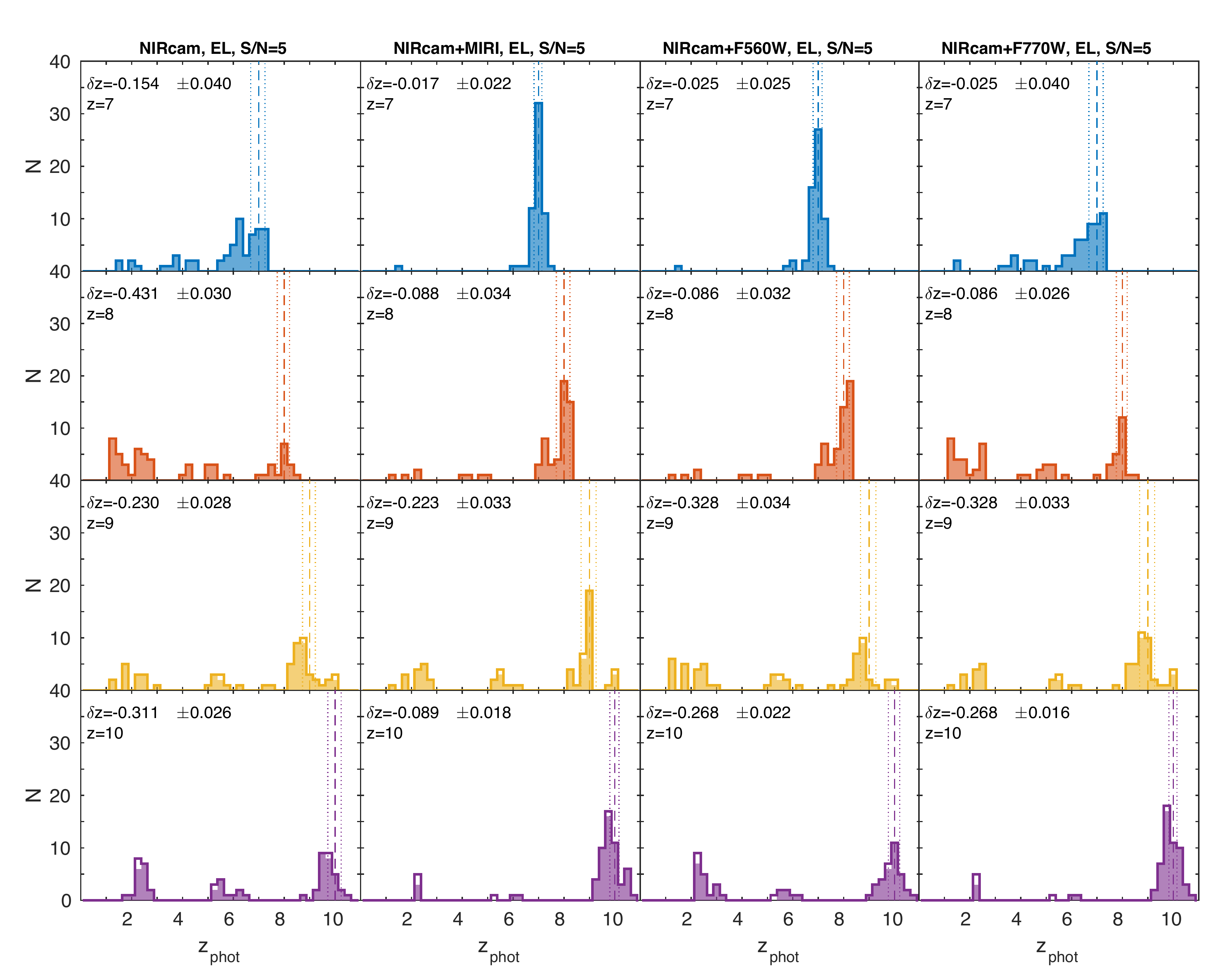}}
\caption{The same as Fig. \ref{fig:z7_YggSN5}, but only for galaxies with emission lines. \label{fig:z7_YggSN5el}}
\end{figure*}

\begin{figure*}[ht!]
\center{
\includegraphics[width=1\linewidth, keepaspectratio]{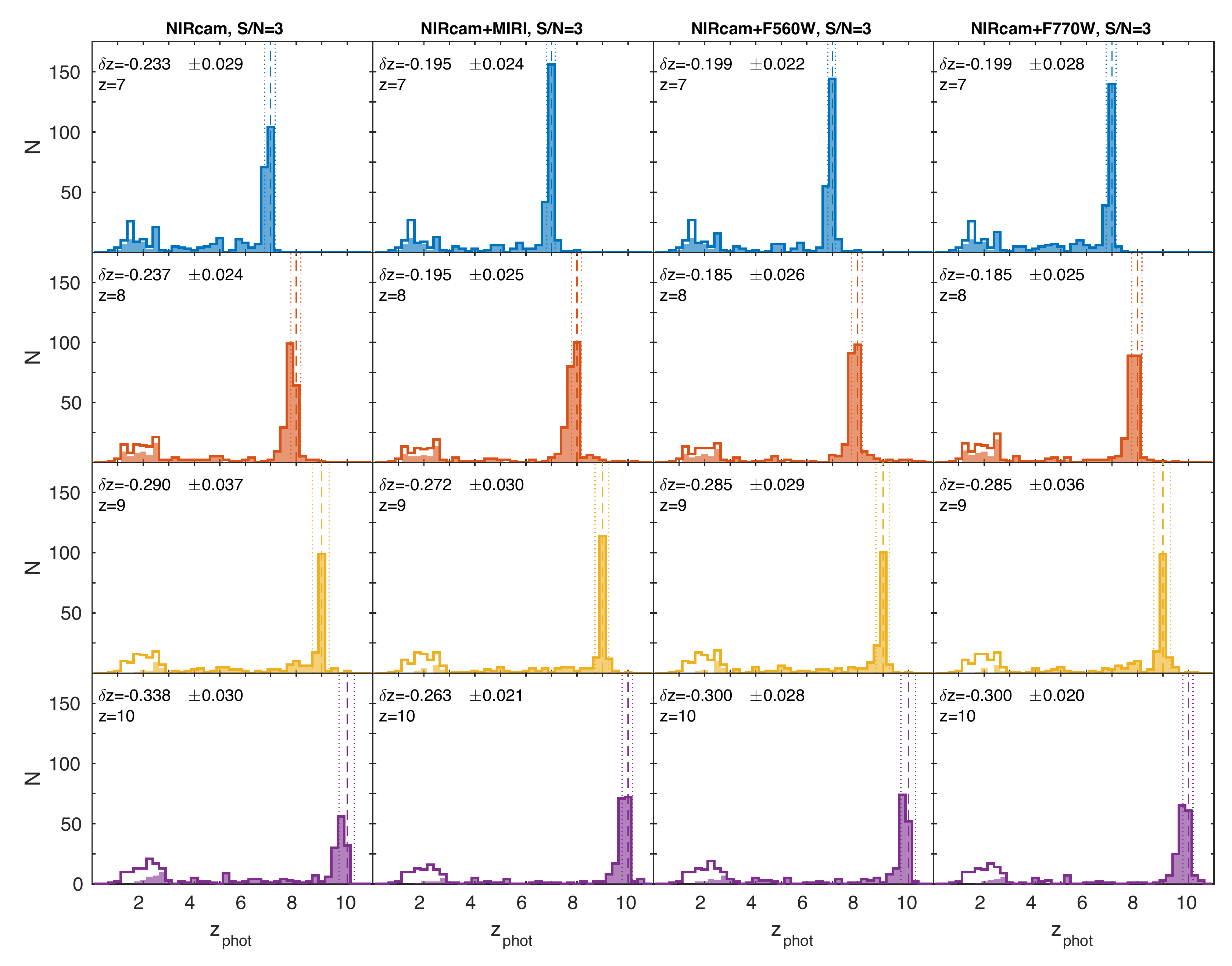}}
\caption{The same as Fig. \ref{fig:z7_YggSN10}, but for $S/N=3$. The shaded area within each histogram corresponds to objects with more than 4 detections in total in the NIRCam and MIRI bands, while empty areas correspond to sources with less than 4-band detections. \label{fig:z7_YggSN3}}
\end{figure*}

In the case of $S/N=3$ at F150W (Fig.~\ref{fig:z7_StN3}), assuming identical integration times in all NIRCam filters implies that some galaxies are undetected at the shortest wavelength NIRCam bands. In this case, having complementary MIRI photometry becomes very important, even with the assumption that we made here, i.e., that the MIRI depth  will be one magnitude shallower than the NIRCam depth at $1.5 \, \rm \mu m$. This assumption is reasonable, as achieving the same depth as in NIRCam will require too long exposure times\footnote{Even achieving a MIRI  F560W depth only one magnitude shallower than the NIRCam depth at F150W will require an integration time $\sim 35$ times longer with MIRI than with NIRCam; see http://jwstetc.stsci.edu/etc/}.  \par 

The result of this poorer-quality photometry is that the output redshift distributions are broader and secondary peaks at lower redshifts become more significant (Fig~\ref{fig:z7_StN3}). The incorporation of MIRI photometry has a significant effect in reducing the overall outlier fraction: it goes down from 10.1\% with NIRCam data only to $3.0\%$ after adding the two MIRI bands (Table~\ref{tab:outliers_z7}). Note that, among the two MIRI filters, F560W appears to have a more important effect in improving the $z_{\rm phot}$ determination of the BC03-SED sources at $z=7-10$. \par

As a summary, we conclude that considering MIRI photometry of one magnitude shallower depth than the NIRCam photometry keeps the fraction of $z_{\rm phot}$ outliers low for the $z\geq7$ galaxies which have a moderate or low $S/N$ in the NIRCam images. This effect is evident for different SED types (with or without emission lines). \par

\subsubsection{Galaxies simulated with Yggdrasil templates at $z=7-10$}

Figures \ref{fig:z7_YggSN10} to \ref{fig:z7_YggSN3} show the distributions of the recovered photometric redshifts for the Sample 3 galaxies simulated with the {\em Yggdrasil} templates at each fixed redshift, obtained by running \textit{LePhare} with different {\em JWST} filter combinations. The outlier percentages for each analysed $S/N$ value are listed in Table \ref{tab:outliers_z7_Ygg}.   {\em Yggdrasil} templates are more complex than the BC03 ones, given that {\em Yggdrasil} automatically incorporates nebular emission lines and continuum emission when galaxy ages are young and the gas covering factor is $f_{\rm cov}>0$, which may be the case of many high-$z$ galaxies. \par

When considering objects with $S/N=10$ at F150W (Figure~\ref{fig:z7_YggSN10}), we see that the NIRCam data alone provides a good photometric redshift recovery ($\sigma=0.01-0.02$). The incorporation of the MIRI bands has little effect on the resulting $\sigma$ values in most cases and  the fraction of catastrophic outliers is only slightly reduced  (see Table~\ref{tab:outliers_z7_Ygg}) \par

As in the case of BC03-SED galaxies, the impact of incorporating MIRI photometry is more evident at $S/N=5$ at F150W (Figure~\ref{fig:z7_YggSN5}).  The overall percentage of catastrophic outliers reduces from $20.4\%$ to $12.1\%$ by adding MIRI photometry in this case (Table~\ref{tab:outliers_z7_Ygg}). \par

If we consider only those galaxies with emission lines in their SEDs (i.e., young and with gas covering factor $f_{\rm cov}=1$), also at $S/N=5$, we see two interesting effects (Figure~\ref{fig:z7_YggSN5el} and Table~\ref{tab:outliers_z7_Ygg}). Firstly, the degeneracies in redshift space are more important than for the general galaxy sample with $S/N=5$. This is manifested in a much larger outlier percentage, namely $\sim 49\%$ when considering the NIRCam data alone. Secondly, the effect of incorporating MIRI photometry becomes more noteworthy: including both MIRI bands reduces the percentage of outliers to $\sim 17\%$. These effects are produced by two reasons: on the one hand, the nebular line emission boosts the MIRI fluxes and, thus, the $S/N$ of the MIRI data.  On the other hand, the multiple emission lines in the \textit{Yggdrasil} templates with nebular emission result in similarly bright observed fluxes in contiguous filters, masking the continuum SED shape and making the SED fitting more difficult (as explained in Section~\ref{sec:z7bc03}), so having a longer baseline is necessary for a proper redshift recovery.\par

For the lowest quality photometry ($S/N=3$; Figure~\ref{fig:z7_YggSN3}), the percentage of catastrophic outliers in the entire sample is $44.4\%$ using only NIRCam photometry.  The advantage of incorporating MIRI data, which also have lower $S/N$, is less obvious in this case, as the percentage of $z_{\rm phot}$ outliers still remains high even with the two MIRI bands ($36.7\%$). A significant fraction of galaxies are detected in only a few (NIRCam and MIRI) bands, so the photometry provides poor constraints for the SED fitting, making degeneracies in redshift space become evident.  \par

Indeed, as it is the case for the BC03-SED galaxies, the catastrophic outliers correspond mainly to sources with detections in four or less {\em JWST} bands (empty regions in the histograms of Figure~\ref{fig:z7_YggSN3}). However, in the case of the  \textit{Yggdrasil}-SED galaxies, the secondary peaks appearing in the output redshift distributions are quite more pronounced than for the BC03 galaxies. These secondary peaks are mainly produced by two kinds of templates that are not represented in our BC03-SED sample: 1) young galaxies with nebular emission. The combination of multiple emission lines and low $S/N$ produces an enhancement in the number of catastrophic outliers; 2) young galaxies with $f_{\rm cov}=0$ (i.e., no emission lines) and the highest extinction adopted to build our mock sample (i.e., $A_V=1$). These galaxies have no analogues either in the BC03-based sample, as we explicitly incorporated the main emission lines in all galaxies with ages lower than the characteristic $\tau$ in the star formation history. \par

As a summary, we conclude that the benefit of considering MIRI photometry along with the NIRCam data for \textit{Yggdrasil}-SED galaxies is mainly obvious for NIRCam sources with moderate $S/N$ values, and particularly helpful in the case of SEDs with nebular emission. At low $S/N$ the advantage is less obvious. This is in contrast to our conclusion for the BC03-SED galaxies, indicating that at low $S/N$ the importance of having MIRI data depends significantly on the galaxy spectral type.  \par

\subsubsection{MIRI magnitudes and NIRCam/MIRI colours}

In this Section we show the MIRI magnitudes and NIRCam/MIRI colours of our simulated galaxies in Sample 3, as this can be a useful reference when planning {\em JWST} observations. In particular, we analyse how the MIRI detections vary with the galaxy SED type. Given our chosen normalisation [F150W]=29~mag, all these galaxies will easily be observed in NIRCam deep surveys (as it takes only $\sim 30$~min integration per pointing to reach [F150W]=29~mag, $3\sigma$ with NIRCam, according to the public {\em JWST}/NIRCam exposure time calculator \footnote{http://jwstetc.stsci.edu/etc/input/nircam/imaging/}). So our aim here is to discuss the fractions of such galaxies that will be detected in typically deep MIRI observations. Of course, these same results only need to be re-scaled to remain valid for other NIRCam magnitudes.\par

\begin{figure}[ht!]
\center{
\includegraphics[width=1\linewidth, keepaspectratio]{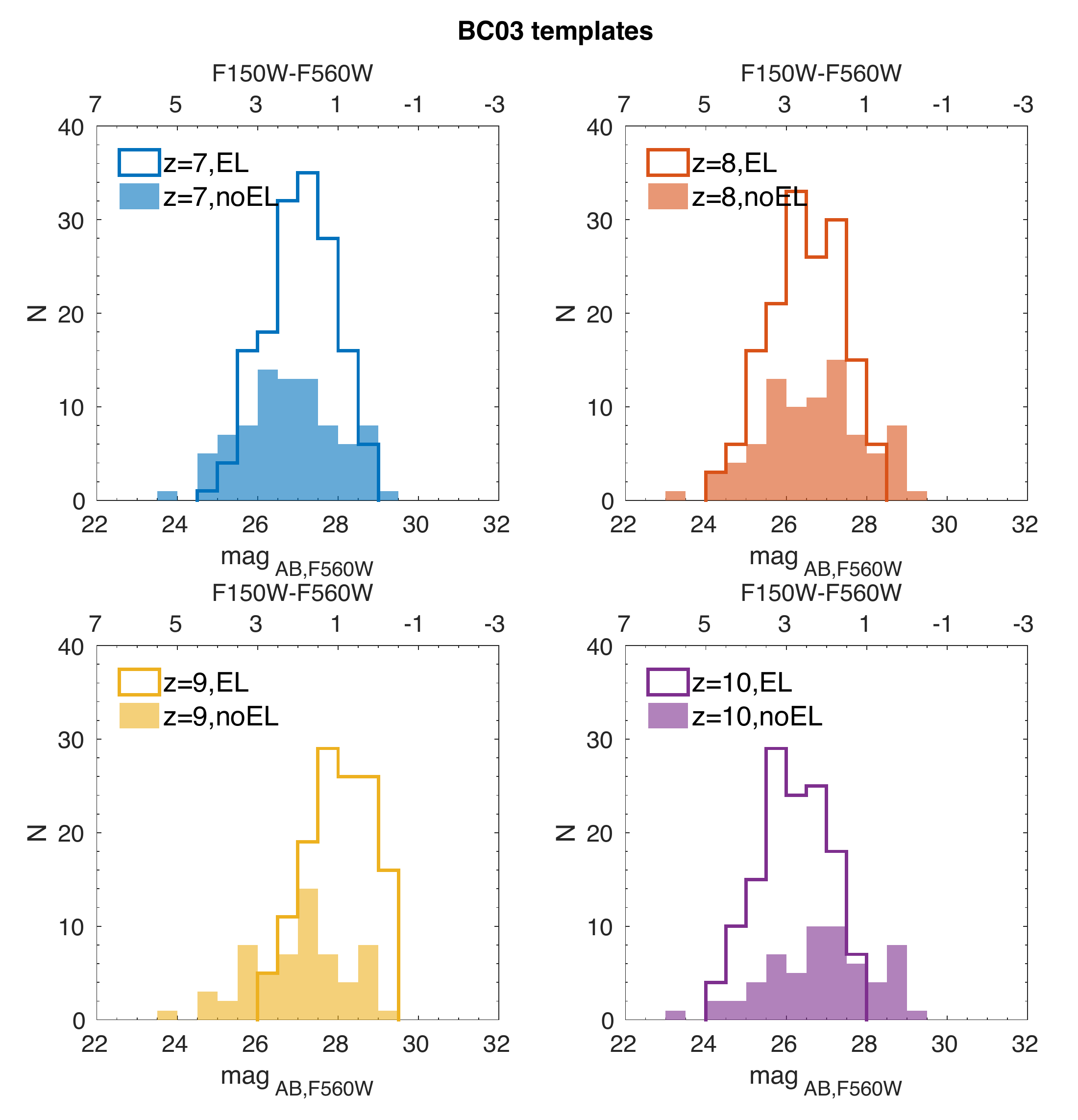}}
\caption{Expected MIRI F560W magnitude distribution for the Sample 3 galaxies simulated with the BC03 models at different redshifts. The cases of galaxies with emission lines (i.e., star-forming galaxies with ages lower than their characteristic SFH $\tau$) and without emission lines are shown in separate histograms.  As the NIRCam F150W magnitude for all these galaxies has been fixed to [F150W]=29~mag, there is a direct correspondence between the MIRI F560W magnitudes and the F150W-F560W colours (top x axis). \label{fig:f560wcol_bc03}}
\end{figure}

\begin{figure}[ht!]
\center{
\includegraphics[width=1\linewidth, keepaspectratio]{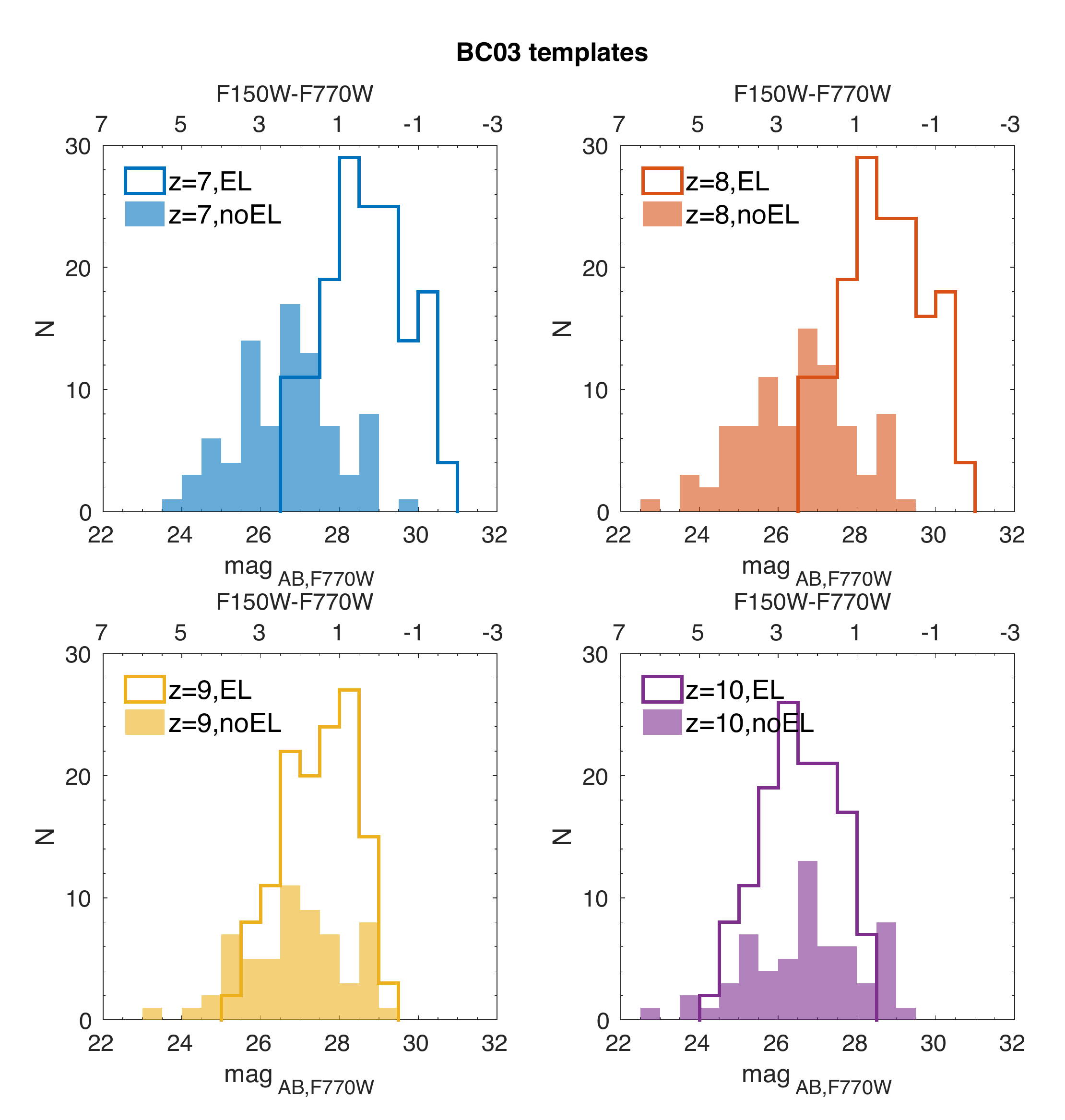}}
\caption{The same as Figure~\ref{fig:f560wcol_bc03}, but for the MIRI F770W magnitudes and F150W-F770W colours. \label{fig:f770wcol_bc03}}
\end{figure}

\begin{figure}[ht!]
\center{
\includegraphics[width=1\linewidth, keepaspectratio]{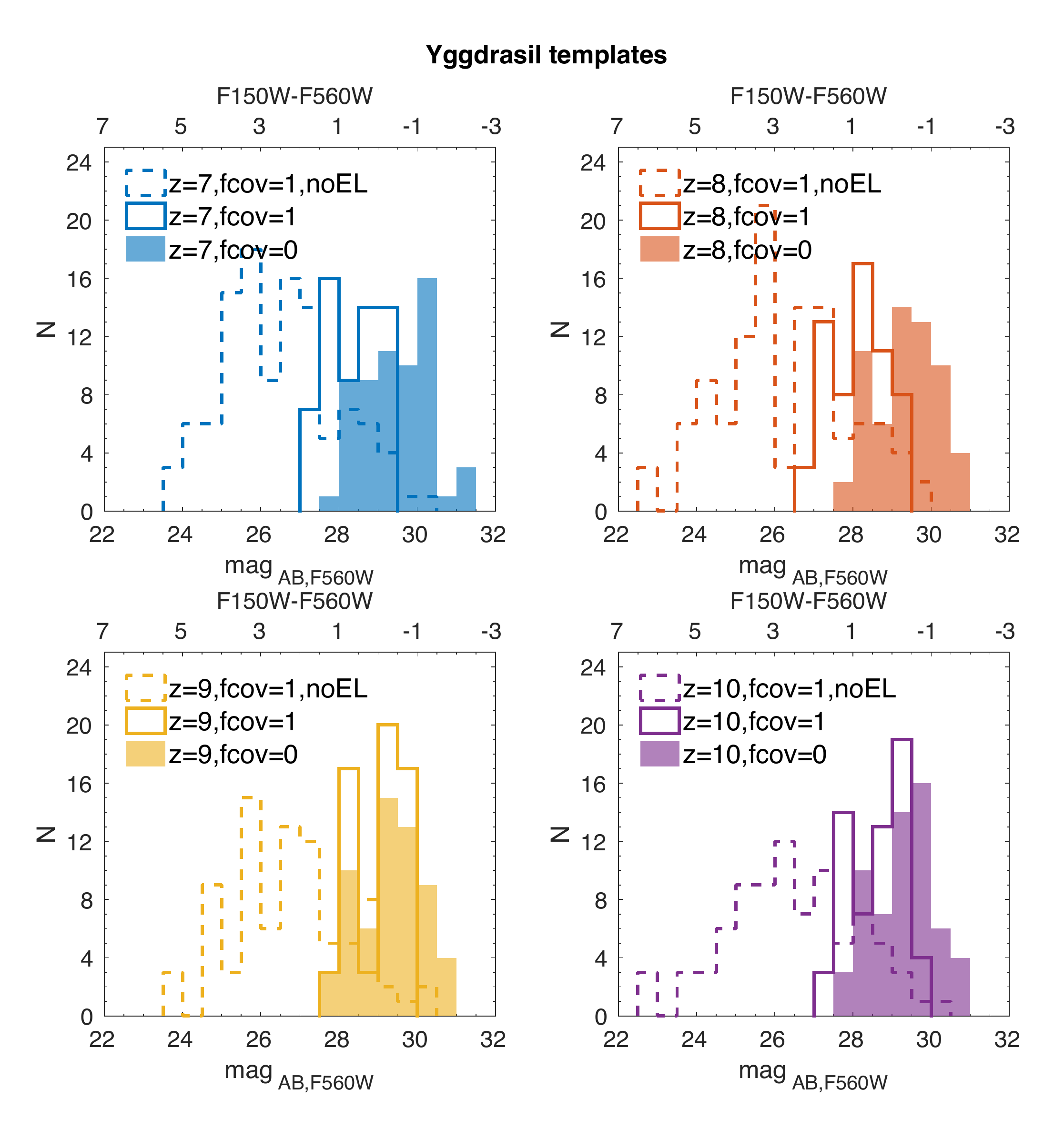}}
\caption{Expected MIRI F560W magnitude distribution for the Sample 3 galaxies simulated with the {\em Yggdrasil} models for galaxies in passive evolution (i.e. ages larger than their constant star formation period, \textit{dashed line}) and galaxies that are still forming stars. The star forming galaxies are also divided in models with covering factors $f_{\rm cov}=0$ (\textit{filled area}) and 1 (\textit{continued line}), as only the latter have emission lines and continuum nebular emission.. Note that the gas covering factor is irrelevant in the cases of galaxies in passive evolution, as no nebular emission is present any more in these galaxies. The top x axis shows the F150W-F560W colours. \label{fig:f560wcol_ygg}}
\end{figure}

\begin{figure}[ht!]
\center{
\includegraphics[width=1\linewidth, keepaspectratio]{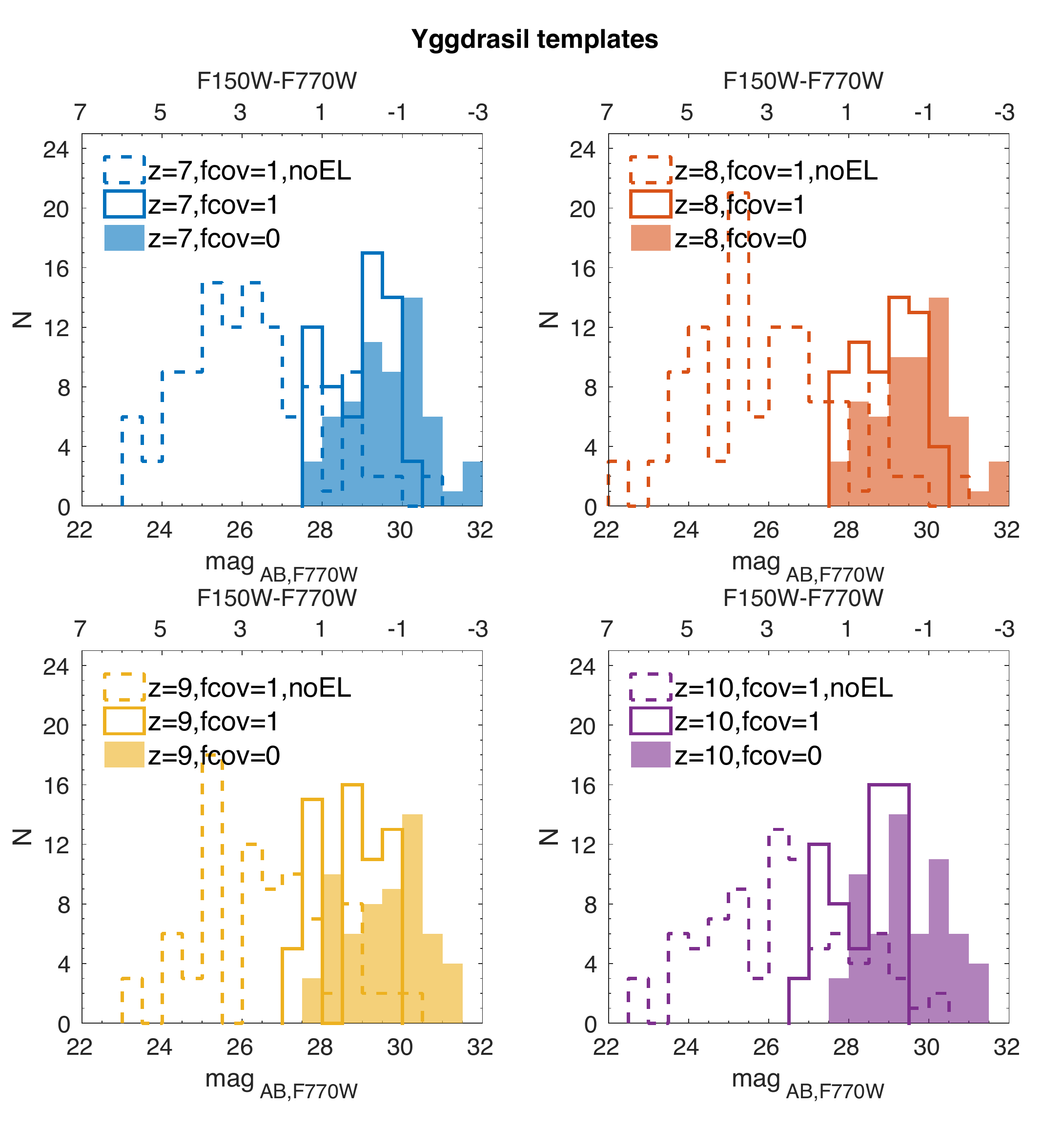}}
\caption{The same as Figure~\ref{fig:f560wcol_ygg}, but for the MIRI F770W magnitudes and F150W-F770W colours. \label{fig:f770wcol_ygg}}
\end{figure}

Figures \ref{fig:f560wcol_bc03} to \ref{fig:f770wcol_ygg} show the expected  MIRI magnitude distributions for our Sample 3 galaxies generated with the BC03 and {\em Yggdrasil} models. As the NIRCam F150W magnitude for all these galaxies has been fixed to [F150W]=29~mag, there is a direct correspondence between the MIRI F560W magnitudes and the F150W-F560W colours, which are also shown in each plot.
For the BC03 galaxies (Fig.~\ref{fig:f560wcol_bc03} and \ref{fig:f770wcol_bc03}), we analysed separately the cases with and without emission lines, which here correspond to galaxies with ages lower and higher than their characteristic SFH decay parameter $\tau$, respectively.  For the \textit{Yggdrasil} galaxies (Fig.~\ref{fig:f560wcol_ygg} and \ref{fig:f770wcol_ygg}), we analysed three cases: galaxies with ages larger that their constant star formation period, which are galaxies in passive evolution; and galaxies which are still forming stars with two different gas covering factors: $f_{\rm cov}=0$ (implying no line or nebular continuum emission) and 1 (with nebular emission). The median values of all distributions are provided in Table~\ref{tab:medmag56} and \ref{tab:medmag77}. \par

The magnitude distributions in Fig.~\ref{fig:f560wcol_bc03} show that the vast majority of BC03 galaxies with [F150W]=29~mag at $z=7, 8$ and 10, and more than a half of those at $z=9$, will be detected by a MIRI survey with a detection limit [F560W]$=28$~mag. This result is valid for galaxies with or without emission lines. All the latter are cases in which a $4000 \, \rm\AA$ break is at least partly developed, so the SEDs are brighter at the MIRI wavelengths than at $1.5 \, \rm \mu m$, particularly at $z=7$ (see Fig.~\ref{fig:sed_z7}). This k-correction effect makes that the median [F560W] value at this redshift is  brighter for galaxies without emission lines, as those with emission lines correspond to younger galaxies with fainter continua which partly cancel the emission line flux enhancement (Table~\ref{tab:medmag56}). At $z=8$ and 10, instead, the combination of the continuum k-correction and the nebular emission results in a net flux boosting in the case of emission-line galaxies. \par

For F770W (Fig.~\ref{fig:f770wcol_bc03}), the flux enhancement due to the emission line presence is only evident at $z=9$ and 10, as the H$\alpha$ emission line enters the F770W wavelength window at these redshifts (Table~\ref{tab:medmag77}). Most young galaxies at $z=7$ and 8, instead,  remain below our reference [F770W]=28~mag detection limit, as their continua are blue and no emission line is present in this filter to help with the detection. So, the F770W filter will only allow us to detect evolved galaxies at $z\geq7$ or young line emitters at $z=9$ and 10 to a [F770W]=28~mag cut. \par

The vast majority of the {\em Yggdrasil}-SED galaxies with [F150W]=29~mag that finished forming stars at $z\geq7$ will have mag$<28$ in both MIRI bands (Fig.~\ref{fig:f560wcol_ygg} and \ref{fig:f770wcol_ygg}, and Tables~\ref{tab:medmag56} and \ref{tab:medmag77}). This is the result of the same k-correction effect discussed before. \par

For the \textit{Yggdrasil} star-forming galaxies, there is a clear enhancement of the MIRI fluxes produced by the nebular lines (compare the cases with $f_{\rm cov}=0$  and 1 in Fig.~\ref{fig:f560wcol_ygg} and \ref{fig:f770wcol_ygg} and the corresponding Tables). In this case, the flux enhacement is more obvious that for the BC03 templates because here we are comparing only young galaxies with and without lines (segregated by their different gas covering factors), so all their continua are similarly blue. In both cases, however, the median MIRI magnitude values are $>28$~mag. This implies that a MIRI galaxy survey to our reference depth will only detect a minority of star-forming galaxies with [F150W]=29~mag at $z\geq7$ if their SEDs are similar to the \textit{Yggdrasil} templates. Note that these galaxies are considerably brighter in F560W if they are at $z=7$ or 8, while slightly brighter in F770W if at $z=9$ or 10. \par

Note that, conversely, our colour analysis shows that detecting more than a half of red MIRI galaxies at $z\geq7$ will require NIRCam data at least three magnitudes deeper. This is the case both for `passive galaxies' (galaxies which finished forming stars)  and for galaxies which are red simply because of the presence of prominent emission lines. \par

So, as a summary, `old' galaxies at $z\geq7$ will easily be detected with MIRI in reasonably deep surveys, independently of their galaxy type. For young galaxies, the situation is more complex, as it depends on to which extent  line emission can compensate for the very faint continua which is characteristic in these sources.  F560W will provide more detections than F770W if maps of the same depth are obtained in both passbands. \par

\begin{deluxetable*}{cccccc}[ht!]
\tablecaption{Median MIRI [F560W] magnitudes for our Sample 3 galaxies at $z=7-10$ (all normalised at [F150W]=29~mag).  \label{tab:medmag56}}
\tablecolumns{6}
\tablewidth{0pt}
\tablehead{
\colhead{Redshift} & 
\colhead{BC03} & 
\colhead{BC03} & 
\colhead{\textit{Yggdrasil}} & 
\colhead{\textit{Yggdrasil}} &
\colhead{\textit{Yggdrasil}}   \\
& No emiss. lines & With emiss. lines & Passive evolution & Star formation,  $f_{\rm cov}=0$ & Star formation, $f_{\rm cov}=1$
}
\startdata
7  & 26.86 & 27.08  & 26.28 & 29.50  & 28.47  \\
8  & 26.77 & 26.49  & 25.96 & 29.43 & 28.32  \\
9  &  27.12 & 28.08 & 26.66  & 29.35  & 29.15  \\
10 & 26.96 & 26.17  & 26.44 &  29.27  & 28.72 \\  
\enddata
\end{deluxetable*}

\begin{deluxetable*}{cccccc}[ht!]
\tablecaption{Median MIRI [F770W] magnitudes for our Sample 3 galaxies at $z=7-10$ (all normalised at [F150W]=29~mag).  \label{tab:medmag77}}
\tablecolumns{5}
\tablewidth{0pt}
\tablehead{
\colhead{Redshift} & 
\colhead{BC03} & 
\colhead{BC03} & 
\colhead{\textit{Yggdrasil}} & 
\colhead{\textit{Yggdrasil}} &
\colhead{\textit{Yggdrasil}}   \\
& No emiss. lines & With emiss. lines & Passive evolution & Star formation,  $f_{\rm cov}=0$ & Star formation, $f_{\rm cov}=1$
}
\startdata
7  &  26.70 & 28.64  & 26.10  & 29.64  & 29.05  \\
8  &  26.59 & 28.68  & 25.80  & 29.62  & 29.13 \\
9  &  26.98 & 27.52  & 26.44  & 29.58  & 28.87  \\
10 &  26.79 & 26.51  & 26.24  & 29.47  & 28.53 \\  
\enddata
\end{deluxetable*}

\section{Summary and conclusions}\label{sec:conclusions}

In this work we have tested the impact of having data in different {\em JWST} filter combinations on deriving photometric redshifts for galaxies at different redshifts. We considered the 8 NIRCam broad bands and the two shortest wavelength MIRI bands (F560W and F770W), which are the most sensitive ones and, thus, those which will be preferred for the study of high-$z$ galaxies. In addition,  we also investigated the effect of having (or not) ancillary photometry from {\em HST} and ground-based telescopes at wavelengths shorter than those to be observed with {\em JWST} (i.e., $\lambda < 0.6 \, \rm \mu m$). \par

We performed our tests on three galaxy samples with known input redshifts: a sample of 2422 galaxies with spectroscopic redshifts $z=0-6$ (Sample 1); a sample of 1375 galaxies with CANDELS consensus photometric redshifts at $z=4-7$ (Sample 2); and a sample of 2124 mock galaxies at $z=7-10$ (Sample 3), whose SEDs have been simulated with two template libraries, namely BC03 with manual addition of the main emission lines for star-forming galaxies (considered as those with age lower than the characteristic $\tau$ of the star formation history), and the \textit{Yggdrasil} library.    Besides, for Sample 3 we also explicitly investigated the effect of having photometry with different $S/N$ values on the ability to recover the galaxy redshifts. In Samples 1 and 2, more than 90\% of galaxies have F150W photometry with $S/N>5$, so the test results based on these samples are not affected by the photometric quality. \par

These three samples altogether have allowed us to assess different effects in the photometric redshift estimation, such as the presence of high-$z$ contaminants and the leakage of high-$z$ sources towards low $z$. Note that our three samples are not meant to emulate a real galaxy population selected from a {\em JWST} NIRCam or MIRI blank field, but rather sample the typical redshifts and SED types of the galaxies that will be observed in these surveys, in order to test potential problems in the $z_{\rm phot}$ derivation. \par

Our main results are:

$\bullet$ {\em The NIRCam broad bands alone are not sufficient to obtain good-quality photometric redshift estimates at low and intermediate redshifts ($z<7$)}. For Sample 1, which is optimised to study the redshift range $z=0-4$, the use of NIRCam data alone results in a high photometric redshift dispersion and outlier percentage ($>10\%$). In Sample 2, which is particular suited to study the leakage of $z=4-7$ sources to low $z$ due to redshift failure,  we found that $\sim 20\%$ ($\sim 9\%$) of the input sources with $z=4-5$ ($z=5-7$) are misidentified as lower $z$ galaxies. These effects are smaller at $z>6$, as the Lyman break is shifted into the NIRCam bands. 

$\bullet$ {\em Having photometry in the HST F435W and F606W bands, and if possible also in the ground-based $U$ band, is very important to have the fraction of low-$z$ contaminants  under control}. This photometry is important to constrain the Lyman break shift and, thus, the redshifts of low-$z$ sources. Indeed, in Sample 1 the percentage of low-$z$ outliers changes from $>10\%$ to $\sim 7\%$, and the normalised redshift difference distribution r.m.s. reduces in 40\%. In Sample 2, the percentage of sources leaking from $z=4-5$ ($z=5-7$) to lower $z$ reduces from $\sim 20\%$ ($\sim 9\%$) to $\sim 2\%$ ($\sim 8\%$). At $z=5-7$, the reduction in the percentage of leaking sources is exclusively due to the incorporation of the {\em HST} bands. The additional incorporation of $U$-band data has no effect because the Lyman break shifts inside the {\em HST} wavelength range at these redshifts.

$\bullet$ {\em MIRI F560W and F770W data can help mitigate the absence of photometry at $\lambda < 0.6 \, \rm \mu m$}. In Sample 1, we found that the percentage of contaminants reduces from 41\% (22\%) to  $\sim 11\%$ ($0\%$) at $z=4-5$ ($z=5-7$),  by complementing the NIRCam data only with MIRI data in the photometric input catalogue.  For the leaking sources in Sample 2, the addition of MIRI photometry to the NIRCam data reduces the percentage of outliers from  $\sim 20\%$ ($\sim 9\%$) to $<8\%$ ($<4\%$) at the same redshifts. When both the {\em HST} and MIRI photometry are considered along with the NIRCam photometry, the percentage of leaking sources from $z=4-5$ goes down to $\sim1\%$.

$\bullet$ Among the different photometric configurations tested here, the MIRI F560W and F770W bands are the only ones that can meaningfully complement the NIRCam data in galaxy studies at $z \geq 7$. The importance of the {\em HST} data at $\lambda < 0.6 \, \rm \mu m$ becomes negligible once the Lyman break shifts within the NIRCam bands, except to confirm the expected non-detections.

$\bullet$ For photometry with F150W $S/N=10$, the NIRCam data alone allows for accurate redshift estimates for galaxies at $z\geq 7$ in most cases. We found that  the r.m.s. of the $|z_{\mathrm{phot}}-z_{\mathrm{input}}|/(1+z_{\mathrm{input}})$ distributions for Sample 3 range between $\sigma<0.01$ and 0.02 with NIRCam data alone at all high $z$. The addition of MIRI photometry has a mild effect, being useful to enhance the accuracy of the derived $z_{\rm phot}$.

$\bullet$ For sources with F150W $S/N=5$, the benefit of incorporating MIRI photometry becomes more evident. In our reference case, in which the MIRI depth is one magnitude brighter than the NIRCam F150W depth, the incorporation of MIRI data significantly reduces the fraction of outliers with respect to the case with NIRCam data alone. This improvement is particularly noteworthy in the case of \textit{Yggdrasil}-SED galaxies with emission lines, as the multiple nebular lines coupled with a moderate-quality $S/N$ make the SED fitting more challenging. The MIRI bands help extend the wavelength baseline for the SED fitting, and thus using NIRCam and MIRI data altogether results in a clear redshift recovery improvement.

$\bullet$ At F150W $S/N=3$, the role of MIRI data in improving the redshift determination depends very much on the SED type. As `old' galaxies are red, MIRI fluxes are brighter that NIRCam fluxes, so the incorporation of the MIRI data is of clear benefit for these sources.
For star-forming galaxies, the presence of emission lines in the MIRI bands boosts the MIRI fluxes, but at the same time these galaxies have bluer continua, so the net effect is more variable in these cases.

$\bullet$ At fixed $S/N$,  {\em the presence of multiple emission lines, like in the case of nebular emission, makes the correct redshift identification much more difficult.} Although the emission lines enhance the fluxes, the presence of emission lines in all passbands make the filter-convolved SEDs rather featureless and the SED fitting becomes more complicated.

In addition, we analysed the MIRI magnitudes of our Sample 3 simulated galaxies at $z\geq7$. As we have normalised all the SEDs to [F150W]=29~mag, the F150W-MIRI colours directly correspond to MIRI magnitude values. Analysing this expected MIRI photometry has allowed us to assess what types of high-$z$ galaxies will likely be detected in MIRI deep surveys. For a different F150W normalisation, one simply has to scale these results, as the galaxy colours are fixed for each SED model.

We found that a MIRI survey with a detection limit of [F560W]=$28$~mag should enable the detection of the vast majority of the NIRCam sources with [F150W]=29~mag at $z=7-10$, provided that these sources are at least mildly evolved. The youngest galaxies which are still forming stars (\textit{Yggdrasil}-SED galaxies with ages $<100 \, \rm Myr$) will remain mostly undetected. Although in the case of gas covering factor $f_{\rm cov}=1$ the nebular lines make the MIRI fluxes much brighter than in the case of $f_{\rm cov}=0$, the continua of these young galaxies are very blue, so the resulting MIRI fluxes are not very high even in the case of nebular emission.

At the same MIRI reference depth, the F770W filter will provide less detections than the F560W filter at $z\geq7$. Among the NIRCam sources with [F150W]=29~mag,   evolved galaxies at $z\geq7$ and some line emitters at $z=9$ and 10 will be detected in F770W, but very few line emitters at $z=7$ and 8.

Conversely, our colour analysis shows that detecting the majority of red MIRI galaxies at $z\geq7$ will require NIRCam data at least three magnitudes deeper. This is the case both for `passive galaxies' (galaxies which finished forming stars)  and for galaxies which are red simply because of the presence of main emission lines.

As an overall summary, we argue that the results presented here constitute a useful reference for designing deep imaging surveys with {\em JWST}. We conclude that NIRCam and MIRI will have complementary roles and the optimal observing strategy and filter combination to be adopted will depend on each observing program's science goals. As adding MIRI imaging to the NIRCam observations would significantly enlarge the requested observing times,  observers must clearly assess the need and importance of the former for their specific studies.

MIRI observations with both the F560W and F770W filters are necessary along with the NIRCam data if a correct identification of low and moderate redshift ($z<7$) sources are desired in fields without $\lambda <0.6 \, \rm \mu m$ coverage. Alternatively, a less expensive strategy would be limiting galaxy surveys to fields with ancillary $\lambda <0.6 \, \rm \mu m$ data of matching depth and obtaining only NIRCam data, but the lack of any MIRI imaging would still result in a slightly higher percentage of redshift outliers and a lower overall recovered-redshift precision.

If the main science goal is simply identifying the bulk of the galaxy population at $z=7-10$, proposing an observing program based on ultra-deep NIRCam data alone will be adequate. However, adding MIRI imaging will help improve the identification of red sources and is of utmost importance to recover stellar ages and stellar masses at $z>7$ (Bisigello et al., in prep.). Therefore, observing with both cameras will be necessary for a comprehensive study of galaxy evolution since early cosmic times.

\acknowledgments

This work is partly based on observations taken with the NASA/ESA {\em Hubble Space Telescope}, which is operated by the Association of Universities for Research in Astronomy, Inc., under NASA contract NAS5-26555; with the {\em Spitzer Space Telescope}, which is operated by the Jet Propulsion Laboratory, California Institute of Technology under a contract with NASA; and the Very Large Telescope operated by the European Southern Observatory (ESO) at Cerro Paranal. LB and KIC acknowledge the support of the Nederlandse Onderzoekschool voor de Astronomie (NOVA).  KIC also acknowledges funding from the European Research Council through the award of the Consolidator Grant ID 681627-BUILDUP. OLF acknowledges funding from the European Research Council Advanced Grant ID 268107-EARLY.  PGP-G acknowledges support from the Spanish Government MINECO Grants AYA2012-31277 and AYA2015-70815-ERC. LG acknowledges support from the Spanish Government MINECO Grants AAYA2012-32295. JP acknowledges the UK Science and Technology Facilities Council and the UK Space Agency for their support of the UK's {\em JWST} MIRI development activities. We thank Macarena Garc\'{i}a-Mar\'{\i}n, Alistair Glasse and \'Alvaro Labiano for providing us the most updated versions of the {\em JWST} MIRI filter transmission curves. We also thank an anonymous referee for a careful and constructive report.

\vspace{5mm}




\appendix

\section{Sample 3: analysis of additional redshifts}\label{sec:AppA}
\begin{figure}[ht!]
\center{
\includegraphics[width=0.5\linewidth, keepaspectratio]{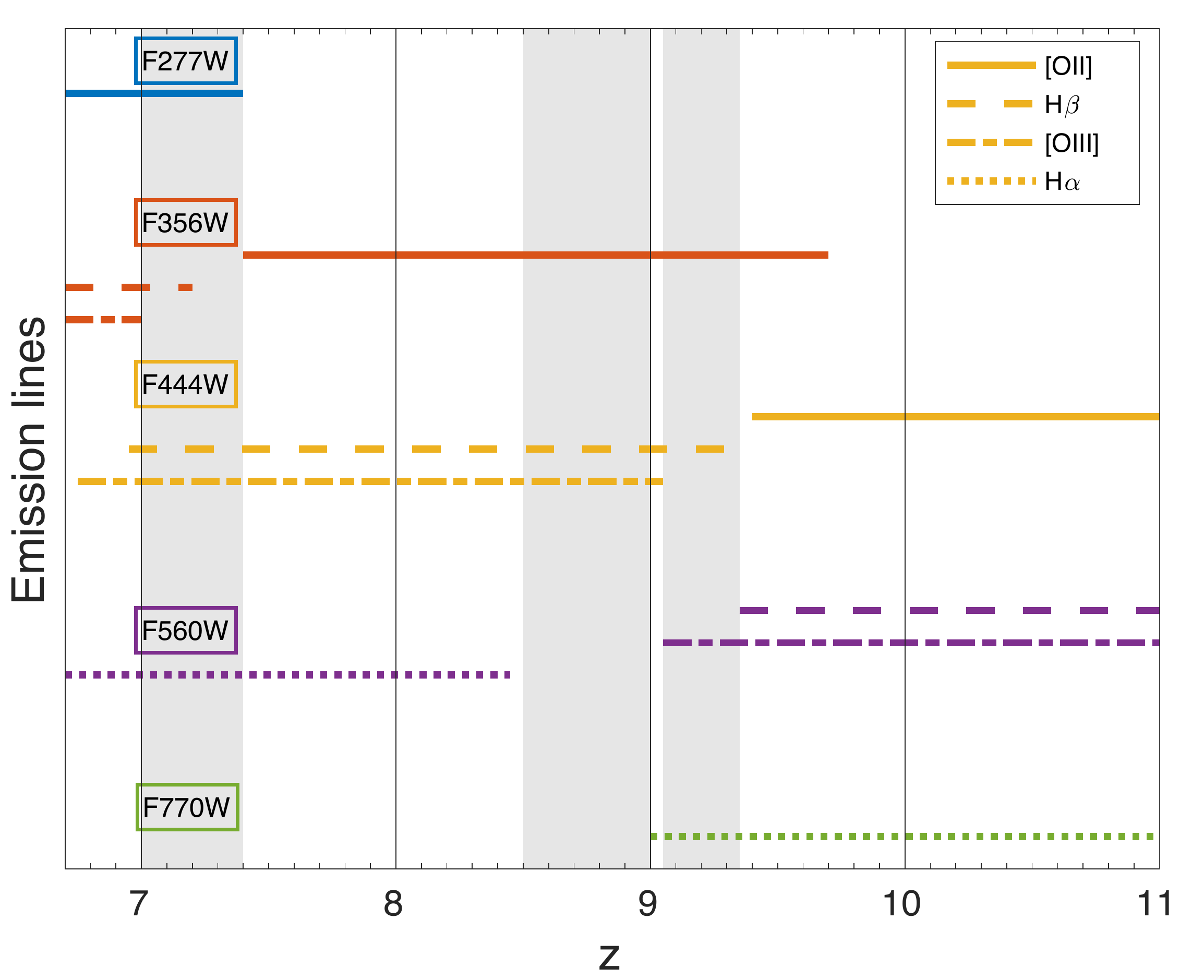}}
\caption{Emission lines present inside NIRCam and MIRI bands at high redshifts. Different colours indicate different bands. \textit{Blue:} F277W;  \textit{red:} F356W; \textit{yellow:} F444W; \textit{purple:} F5602; and \textit{green:} F770W. The shortest wavelength NIRCam bands are not shown because they do not contain emission lines in this redshift range. Different line styles represent different emission lines: [OII] (\textit{continuous line}); H$\beta$ (\textit{dashed line}); [OIII] (\textit{dot-dashed line}); and H$\alpha$ (\textit{dotted line}). Vertical black lines indicate the redshifts considered in sample 3 ($z=7,8,9$ and $10$). These specific redshifts are representative of all $z=7-10$ redshifts, except those in the grey areas, which are instead analysed in this appendix.
\label{fig:el_band}}
\end{figure}

In Section~\ref{sec_sample3descr} we considered four redshifts for our Sample 3 analysis: $z=7$, 8, 9 and 10. These four redshifts are representative of most of the entire redshift range $z=7-10$, except for a few exception that we study here.  For this analysis, we considered only BC03 templates to which we manually added the main emission lines.  We analysed how these emission lines move in and out NIRCam and MIRI filters to identify all possible configurations (Fig.~\ref{fig:el_band}). For each filter, we only took into account the wavelength range where its transmission is above 20$\%$. To cover the redshift ranges  that are unrepresented by our original analysis (corresponding to the shaded areas in Fig.~\ref{fig:el_band}), we re-tested the Sample 3 redshift recovery at four additional redshifts, namely, $z=7.1, 7.3, 8.7$ and $9.2$.\par

\begin{figure}[ht!]
\center{
\includegraphics[width=1\linewidth, keepaspectratio]{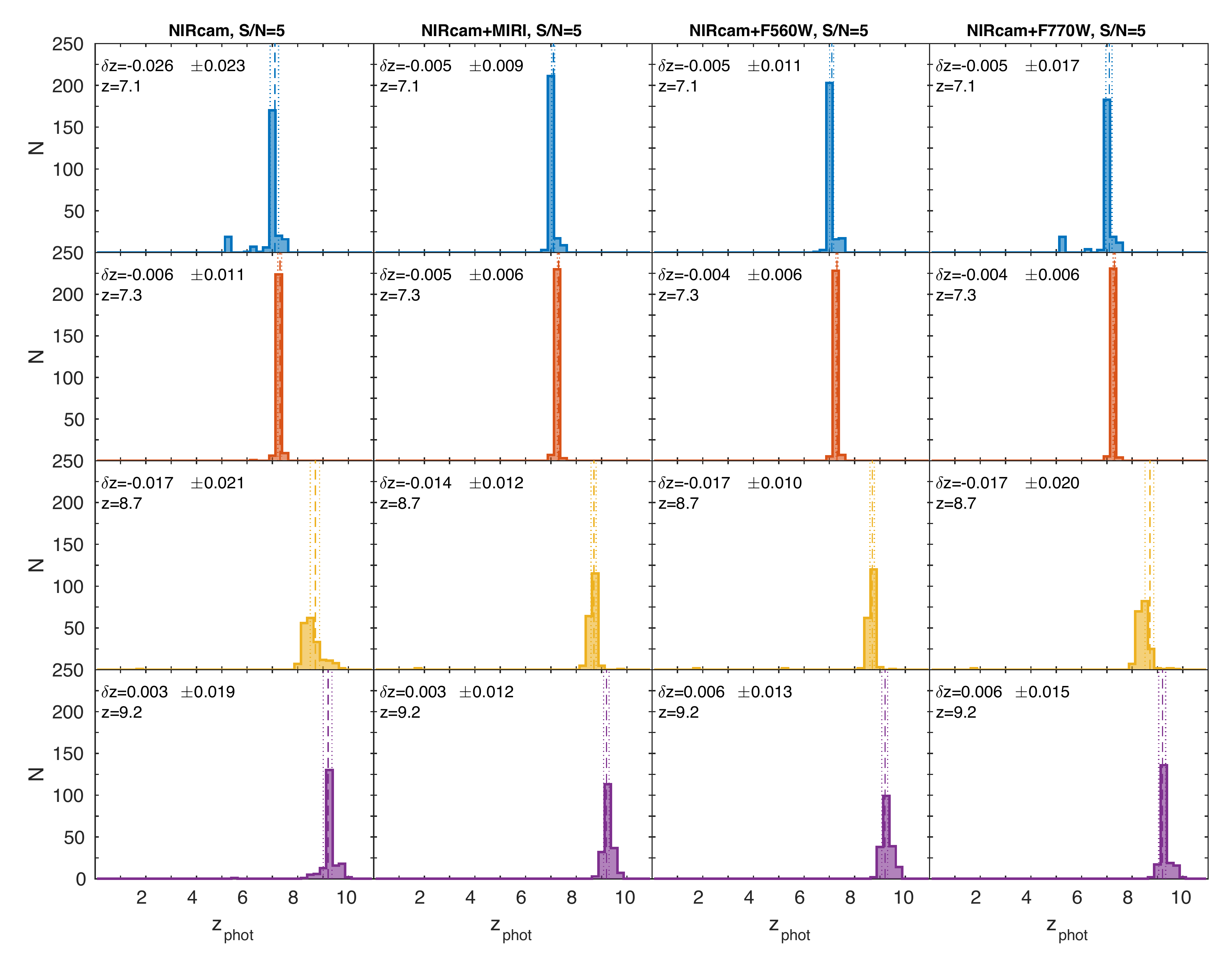}}
\caption{Photometric redshifts obtained for the  BC03 simulated galaxies with (F150W) $S/N=5$ at different fixed redshifts. \textit{From top to bottom}: input redshifts $z=$7.1, 7.3, 8.7 and 9.2.  Photometric redshifts in each column are obtained with different combinations of bands. \textit{From left to right:} 8 NIRCam broad bands; 8 NIRCam broad bands, MIRI F560W and MIRI F770W; 8 NIRCam broad bands and MIRI F560W only; 8 NIRCam broad bands and MIRI F770W only. The vertical lines indicate the $3\sigma$ interval around the mean normalised redshift difference. On the top left of each panel we quote the mean of $(z_{phot}-z_{input})/(1+z_{input})$ and the r.m.s of $|z_{phot}-z_{input}|/(1+z_{input})$.
\label{fig:newz}}
\end{figure}

The results of our new tests are shown in Fig.~\ref{fig:newz} (for $S/N=5$). At $z=7.1$, the main difference respect to the $z=7$ configuration is that the [OIII] line moves out of the F356W NIRCam filter while H$\beta$ is still inside. The results when using NIRCam broad bands alone are very similar to the $z=7$ case, with similar r.m.s. values, 0.023 in this case (compared with 0.025 at $z=7$) and similar distribution shape. The number of outliers decreases from $\sim8\%$ to 0 when adding the F560W band, but it does not change when adding the F770W band. Also the r.m.s. improves more when adding the F560W filter rather than with the F770W one.

At $z=7.3$ the main difference respect to the $z=7$ configuration and the previous one is that there are no emission lines inside the F356W band. When using NIRCam bands alone, it is already possible to obtain a good redshift estimation with totally no outliers and a small r.m.s.. This is because the absence of emission line inside the F356W band creates a strong contrast with the closest bands, allowing the code to properly identify the main SED features. Moreover, this contrast happens in a narrow redshift range, confining the range of possible redshift solutions and producing a very small r.m.s. value. This is the most favourable configuration for the NIRCam bands among all considered redshifts. By adding the MIRI bands there is only a slight improvement to the already good redshift estimation.

The case of $z=8.7$ is representative of the redshift range where no emission lines at all are present in the MIRI bands, while [OIII], H$\beta$ and [OII] are inside the NIRCam wavelength range. The main difference between this case and the $z=8$ one is that the flux observed in the F115W band, where the Lyman break falls, is fainter and more often is below the 2$\sigma$ detection limit. When using the NIRCam bands only the derived distribution is broad, but there are almost no outliers (0.5$\%$). When adding the MIRI bands, and in particular the F560W band, the r.m.s. becomes about a half.

Finally, the $z=9.2$  case corresponds to a redshift in which [OIII] is just outside the F444W band, but H$\beta$ is still inside, while the opposite situation happens in the F560W MIRI band. The redshift estimation with NIRCam bands alone is generally good, with an r.m.s. of 0.019 and only one outlier. When adding the MIRI bands, there is a slight improvement in the redshift estimation (r.m.s.$\sigma$=0.012) and  no outliers.

The cases presented here are slightly different from the cases presented in the main body of the paper, but do not lead to any significantly different conclusion. Considering all eight analysed redshifts and $S/N=5$, the redshift estimation by using only NIRCam broad bands alone is particularly good at $z\sim7.3$ and $8$, but is the worst at $z=8.7$ and $9$. Adding MIRI bands, the redshift estimation improves at all considered redshifts, and, in the majority of cases, F560W decreases the r.m.s. a bit more than F770W.

\section{Sample 3: an alternative observational strategy}\label{sec:AppB}

In Sample 3 we derived fluxes for all NIRCam broad bands assuming the same integration time for all of them. However, the sensitivity of NIRCam peaks around 2 $\mu$m, so different integration times could be considered in order to compensate for different sensitivities. Therefore, we tested this alternative scenario by considering observations of the same depth for all NIRCam bands, i.e., the F150W band depth, and we re-derived photometric redshifts for all galaxies in Sample 3. In particular, this method implies deeper observations for the less sensitive bands, i.e. F070W, F090W, F115W and F444W, respect to the method considered before, but shallower observations for the most sensitive bands, i.e. F200W and F277W. We did this test for both BC03 and Yggdrasil templates and for all three $S/N$ values considered before (3,5 and 10). \par
\begin{figure}[ht!]
\center{
\includegraphics[width=1\linewidth, keepaspectratio]{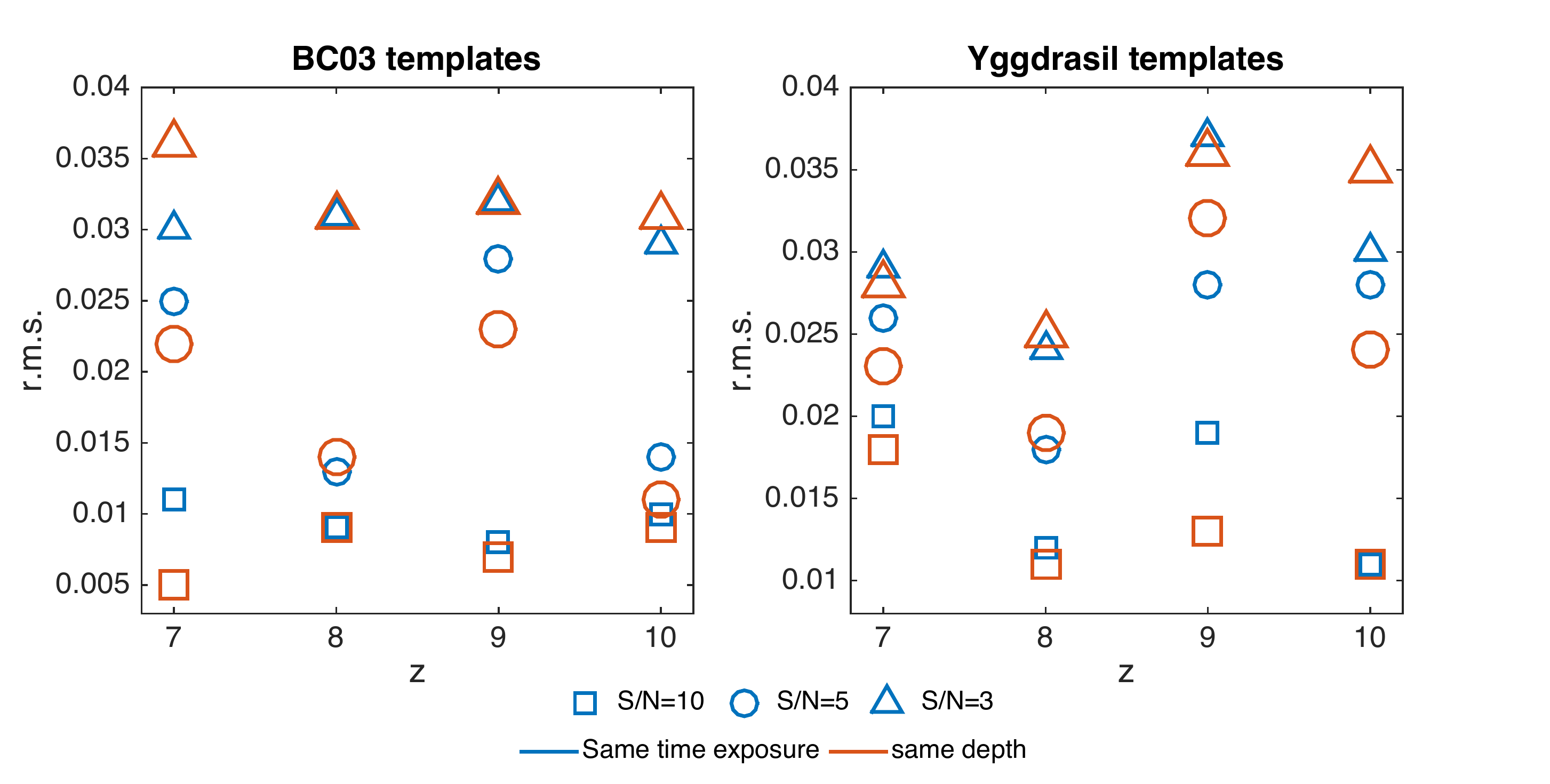}}
\caption{Redshifts vs. r.m.s values of $|z_{phot}-z_{input}|/(1+z_{input})$ derived with the NIRCam broad bands considering the same integration time (\textit{blue symbols}) or the same observational depth (\textit{red symbols}). Different symbols correspond to different $S/N$ values: $S/N=10$ (squares), 5 (circles) and 3 (triangles). {\em Left}: BC03 templates; {\em right}: \textit{Yggdrasil} templates.\label{fig:rms_comparison}}
\end{figure}

We compared the mean and r.m.s. of the normalised redshift difference $|z_{phot}-z_{fiduc.}|/(1+z_{fiduc.})$ for the current case and that where we assumed the same integration time in all NIRCam filters. Our results are shown in Table~\ref{tab:sameStN}, while the r.m.s. values are also shown in Figure~\ref{fig:rms_comparison}. At $S/N=10$, the derived results considering the same depth are slightly better or similar than the one derived with the same integration times. On the other hand, at $S/N=3$, results are similar or slightly worse. At $S/N=5$ the results are mixed, depending on the type of template and redshift considered. These results reveal a different importance of deep observations with the most sensitive and less sensitive NIRCam bands. In particular, deep observations with the less sensitive bands seem to improve results generally when the $S/N$ is high, while deep observations with the most sensitive bands are more crucial at low $S/N$ values. It is worth noticing that, despite the fact that reaching the same depth in all NIRCam filters improves the results derived using NIRCam bands alone, these results are still worse that those derived using NIRCam and MIRI bands together.

\begin{deluxetable}{c|ccc|ccc}[ht!]
\tablecaption{Mean of $(z_{phot}-z_{input})/(1+z_{input})$ and r.m.s of $|z_{phot}-z_{input}|/(1+z_{input})$ for two different observational strategies: the same integration time and the same depth in all NIRCam filters. Photometric redshifts are derived considering only NIRCam bands for four redshifts ($z=7,8,9$ and $10$) and both  BC03 and {\em Yggdrasill} templates.   \label{tab:sameStN}}
\tablecolumns{7}
\tablewidth{0pt}
\tablehead{
\colhead{} &
\multicolumn{3}{@{}c}{{}Same integration time} &
\multicolumn{3}{@{}c}{{}Same observational depth} \\
\colhead{} &
\colhead{$S/N=10$} &
\colhead{$S/N=5 $} &
\colhead{$S/N=3 $} &
\colhead{$S/N=10$} &
\colhead{$S/N=5 $} &
\colhead{$S/N=3 $} 
}
\startdata
z=7 (BC03) & -0.009$\pm$0.011  & -0.030$\pm$0.025 & -0.057$\pm$0.030 &  -0.007$\pm$0.005 & -0.030$\pm$0.022 & -0.070$\pm$0.036\\
z=8 (BC03) & -0.021$\pm$0.009 & -0.031$\pm$0.013 & -0.018$\pm$0.031 &  -0.017$\pm$0.009 & -0.023$\pm$0.014 & -0.025$\pm$0.031\\
z=9 (BC03) & -0.003$\pm$0.008  & 0.001$\pm$0.028 & -0.018$\pm$0.032 &  -0.003$\pm$0.007 & 0.003$\pm$0.023 & -0.011$\pm$0.032\\
z=10 (BC03) & -0.002$\pm$0.010  & -0.006$\pm$0.014 & -0.083$\pm$0.029 &  -0.002$\pm$0.009 & -0.080$\pm$0.011 & -0.042$\pm$0.031\\
z=7 (Ygg) & 0.001$\pm$0.020  & -0.067$\pm$0.026 & -0.233$\pm$0.029 &  -0.005$\pm$0.018 & -0.036$\pm$0.023 & -0.209$\pm$0.028\\
z=8 (Ygg)& -0.022$\pm$0.012  & -0.209$\pm$0.018 & -0.237$\pm$0.024 &  -0.018$\pm$0.011 & -0.164$\pm$0.019 & -0.246$\pm$0.025\\
z=9 (Ygg)& 0.005$\pm$0.019 & -0.084$\pm$0.028 & -0.290$\pm$0.037 &  -0.001$\pm$0.013 & -0.031$\pm$0.032 & -0.244$\pm$0.036\\
z=10 (Ygg)& -0.013$\pm$0.011  & -0.122$\pm$0.028 & -0.338$\pm$0.030 &  -0.015$\pm$0.011 & -0.070$\pm$0.024 & -0.307$\pm$0.035\\
\enddata
\end{deluxetable}

\end{document}